\documentclass[%
 preprint, linenumbers,
 amsmath,amssymb,
 aps, physrev,
]{revtex4-2}

\usepackage{graphicx}
\usepackage{epstopdf}
\usepackage{dcolumn}
\usepackage{bm}
\usepackage{hyperref}
\usepackage{newtxtext}
\usepackage{newtxmath}
\usepackage{ar}
\usepackage{subcaption}
\hypersetup{
    colorlinks = true,
    urlcolor   = blue,
    citecolor  = black,
}

\newcommand{\RomanNumeralCaps}[1]
\linenumbers
\DeclareMathOperator{\sign}{sign}

\begin{document}

\nolinenumbers


\title{\textbf{Analytical Solutions of the Minimal Nonlinear Equation for the Yaw Response of Tail Fins and Wind Vanes.} 
}%

\author{Mohamed M. Hammam}
 \affiliation{Department of Mechanical Power Engineering, Faculty of Engineering, Port Said University, Port Said, Egypt.}
 \email{Contact author: m.hammam@eng.psu.edu.eg}
\author{David H. Wood}%
 \affiliation{Department of Mechanical and Manufacturing Engineering, University of Calgary, 2500 University Dr., Calgary,Alberta T2N 1N4, Canada.
}%

\date{\today}

\begin{abstract}
Analytical solutions for the yaw response of tail fins for small wind turbines, and wind vanes for wind direction measurement, are derived for any planform  and any release angle $\gamma_0$.  This  extends current linear models limited to small $|\gamma_0|$ and low aspect ratio planforms. 
The  equation studied here is the minimal form of the  general second order equation for the yaw angle, $\gamma$, derived by \citet{HW23}.  The nonlinear damping is controlled by a small parameter that depends on the vortex flow coefficient, $K_v$, which is absent from all linear models. The minimal equation is analysed using perturbation techniques. A truncated series solution from the  Krylov-Bogoliubov-Mitropolskii  averaging method  compares favourably with a numerical solution apart from some small deviations at large time.  Another form of averaging due to \citet{BEECHAM1971} yields a compact solution in terms of  the rate of amplitude decay,  and the rate of change of phase angle. This  allows the identification of an equivalent linear system with equivalent frequency and damping ratio. Two limiting analytic solutions for small and large $|\gamma_0|$ are obtained. The former is used to identify the model parameters  from experimental data.  Both approximate solutions showed that high $K_v$ is important for fast decay of  yaw amplitude for tail fins at high $|\gamma_0|$.  High aspect ratios for wind vanes would reduce the nonlinearity to minimize yaw error. Linear response that is independent of $K_v$ occurs whenever $\sin{(\pi \gamma_0)\approx \pi \gamma_0}$. Further, the low angle analytical solution allows an exact identification of the nonlinearity which could be used to extend the modelling of wind vanes to high $\gamma$. 
\end{abstract}

\maketitle


\section{\label{sec:level1}Introduction}

Many small horizontal-axis wind turbines use a tail fin to align the rotor with the wind, e.g. \citet{Wood2011} and \citet{Brad16}.  The unsteady aerodynamics of tail fins is very similar to that of wind vanes used to measure wind direction, e.g. \citet{Kris94}, \citet{Hristov00} and \citet{Ker16}.  All these references used a second order, linear equation to describe unsteady yaw so the response can be characterized by the damping ratio and natural frequency which both depend on the fin or vane planform, e.g. \citet{Singh12} and  \citet{Ker16}.  In addition, the frequency is proportional to the wind speed, $U$, \citet{kedr23}.  The linear solution is limited to small magnitudes of the yaw angle, $\gamma$, and low aspect ratio, $\AR$, defined as $\AR= b_0^2/A$, where $b_0$ is the maximum span and $A$ is the planform area.

The low-$U$  behaviour of fins and vanes can be more complex and problematical than at higher speed.  For example, the acceleration of the blades of a small turbine from rest at around 3 m/s, the typical ``cut-in'' wind speed for the commencement of power production, may involve large yaw angles and long starting times, \citet{Wood2011}.  In addition, the frictional resistive torque in the bearings of fins and vanes may  be significant at low wind speed.  This paper describes work aimed at extending the yaw modelling to the nonlinear range to account for large yaw angles and to allow the modelling of friction.  The theoretical extension  is described by \citet{HW23}. \citet{kedr23} summarize the full nonlinear equations and compare their behaviour to wind tunnel experiments which include the ``Test s'' or TCs that will be the subject of our analysis.  Model tail fins without a rotor or nacelle were mounted in a wind tunnel at an initial yaw angle, $\gamma_0$, and released.  Their subsequent oscillatory  motion was recorded and the results compared to the theoretical predictions for a range of planforms, size, and $\AR$.  The lack of rotor and nacelle make the studies equally applicable to tail fins and wind vanes so that  ``fin'' is here synonymous with ``vane''.  In nearly all these experiments, $U$ was kept constant.  The theory developed by \citet{HW23} includes a term dependent on the rate of change of $U$ but \citet{kedr23} found that their wind tunnel did not allow a sufficiently rapid change in wind speed to test the extra term.  Hence, we concentrate only on the behaviour for steady $U$. 

Good agreement was found using  model parameters established in prior numerical and experimental tests, and this was improved by the use of system identification of those  parameters that were not known to high accuracy, \citet{kedr25}. A simple analysis in \citet{HW22} and the experiments of \citet{kedr23} showed that nonlinearities in yaw response become important when  $\gamma_0$ exceeds around 45$^\circ$. One of our unexpected, new, and important findings in Sections \ref{sec:epsilon} and \ref{sec:small}, however, is that nonlinearity persists down to very small yaw angles before becoming unimportant. Nevertheless, the full response equation is complicated and requires up to nine model constants, so \citet{HW23} derived a reduced but still nonlinear, equation with fewer unknowns to replace the full equation in appropriate cases, but did not attempt to find its closed form solution. The subsequent work to develop an analytical solution is described herein. \cite{kedr25} found that the reduced model  provided an accurate description of TC response if the model constants are optimized using system identification techniques.  Therefore, we concentrate here on how those constants determine the solution so that fin and vane geometries can be better chosen to optimise the response.

There were several motivations for the present work, beginning with supporting the development of a new tail fin module for  the open-source aeroelastic code for wind turbines OpenFAST which can be downloaded from https://www.nrel.gov/wind/nwtc/openfast.html.  Associated with this is the need to give better guidance for  tail fin design, which currently relies more on rules-of-thumb and aesthetic considerations than on sound aerodynamic principles, \citet{Wood2011} and \citet{Brad16}. Wind vane design is in a similar situation; 
 \citet{Ker16} undertook a design optimization but this was limited by their restriction to linear behaviour.
As yet, no detailed study has been made of the effects of power extraction by the rotor on the flow over the tail fin. The new OpenFAST module assumes a simple reduction in wind speed through the rotor.  It is intended to test the validity of this assumption in wind tunnel and field tests of yawing small turbines.  The effects of the rotor are ignored in this study.

The problem considered here is an example of a nonlinear damped oscillation which has an extensive literature, particularly for the case where the nonlinear  terms are of second order.  Then, the main emphasis is to determine whether the ``damping'' and ``frequency'' change, where the quotation marks emphasize that these quantities are a function of time, and possibly other variables, for a nonlinear system.  The asymptotic expansion of nonlinear equations with a perturbation parameter, results in what is known as ``secular'' terms. These can make the solution   aperiodic, or cause the response to grow in amplitude at large time. The secular terms result from truncation of the series expansion of the equations, to leave a finite number of terms and make the solution manageable, \citet{Mickens1981}. In order to eliminate the secular terms, the amplitude must be a function of time  and the period  a function of the amplitude as will be shown in the solution in Section \ref{sec:solution}. These are basic features that differentiate between the response of linear and nonlinear systems.

Our aim is to find an approximate analytic solution to the  nonlinear model developed by \citet{HW23}. The specific aim is to derive the effective damping ratio and response frequency in terms of the geometric parameters of the tail fin to allow an assessment of their role in providing a desirable yaw alignment.  The parameters include the length of the tail boom, $x_p$, and $\AR$. \citet{HW23} showed that high $\AR$ and large $|\gamma|$ are the main causes of nonlinearity with the latter causing the aerodynamic coefficients to become nonlinear in $\AR$. Here we start with a simplified form of the model in which the ``separation functions'' that control the relative magnitude of the aerodynamic moments, are fixed. We call this the ``minimal model'' and the equation representing it the ``minimal equation''. The impact  of the separation functions on the current solution is discussed in Section \ref{sec:coeff}.

The next section describes the dynamic response equation of \citet{HW23} and the assumptions used to disregard high order terms. We present the equation in a standard form and make its parameters non-dimensional. The section ends with a discussion of the bounding linear solutions for $|\gamma| \to 0$ and $|\gamma| \to \pi/2$ which feature significantly in the subsequent analysis.   Section \ref{sec:solution} describes the use of the averaging method of \citet{Krylov47} and \citet{Bog61}, the Krylov-Bogoliubov-Mitropolskii (KBM) method.  Together with the series representation of the nonlinear term the result is a polynomial expression  of the  amplitude and phase in a form  amenable for analytical solution. 
Identifying and quantifying the nonlinearity parameter and specifying the series truncation errors of the solution are the topics of Section \ref{sec:epsilon}. Section \ref{sec:analytical} describes the analytical solution and an analytical expressions for amplitude and phase  are derived. Validation of the KBM solution by comparison to a numerical solution for the tail fin TCs is presented in Section \ref{sec:calculations}. Section \ref{sec:ARC} uses the more accurate \citet{BEECHAM1971} (BT) averaging  method, to provide a compact form of two important parameters that characterize the response:  the  decay and frequency. This leads to an equivalent linear system whose damping ratio and frequency are quantified. Analysis of the BT models in Section \ref{sec:analysis} includes a derivation for the two limiting cases of low and high $|\gamma_0|$. Section \ref{sec:coeff} discusses the implications of simplifying the reduced model to the minimal form. Conclusions are given in Section \ref{conc}.
\begin{figure}
\centering
\includegraphics[width=\textwidth]{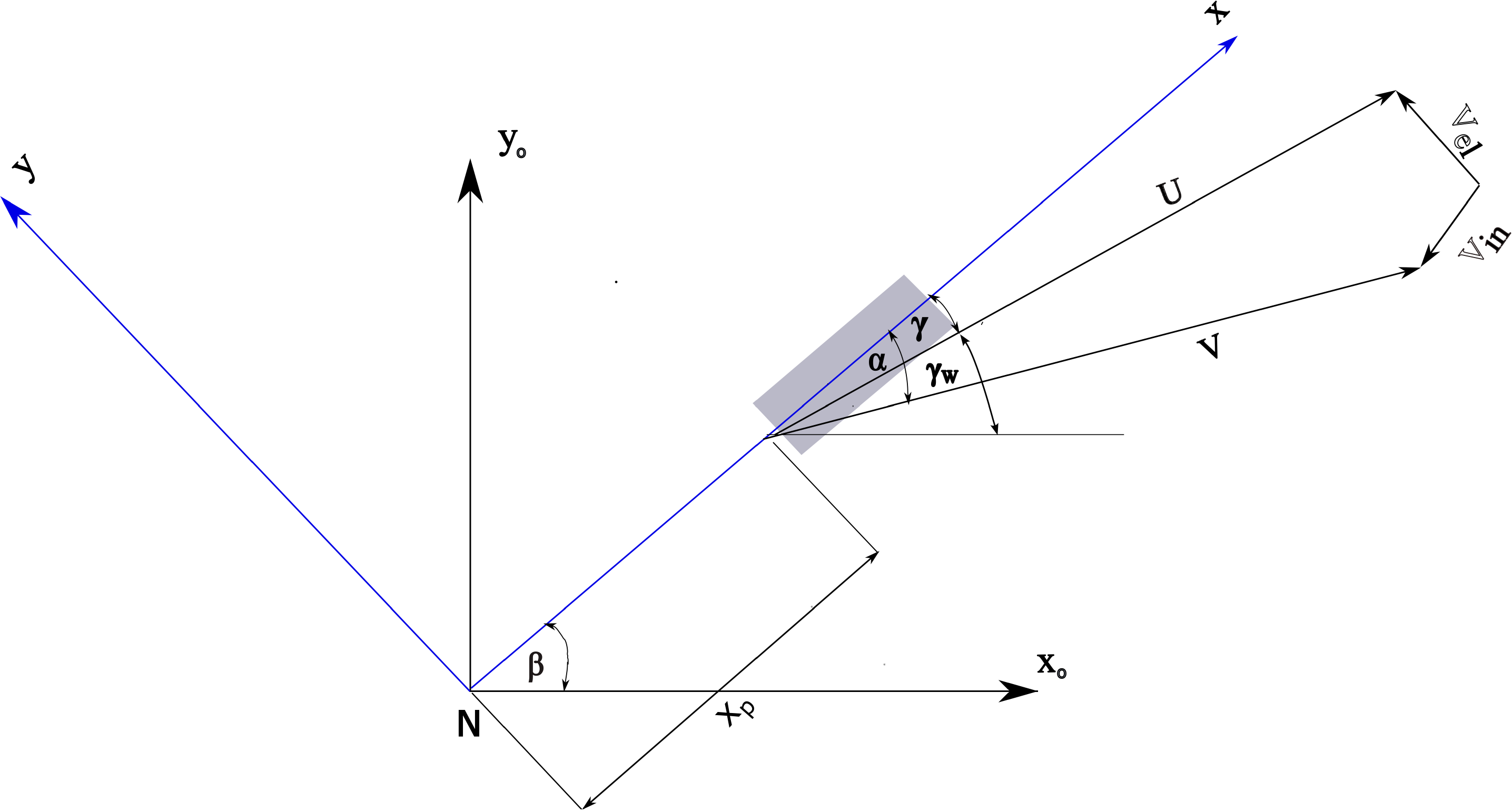}
\caption{Schematic plan view of tail fin motion and coordinate systems. The tail fin of exaggerated thickness is shaded. Figure taken from \citet{HW23}.}
\label{fig:2}
\end{figure}
\begin{figure}
\centering
\includegraphics[width=0.7\textwidth]{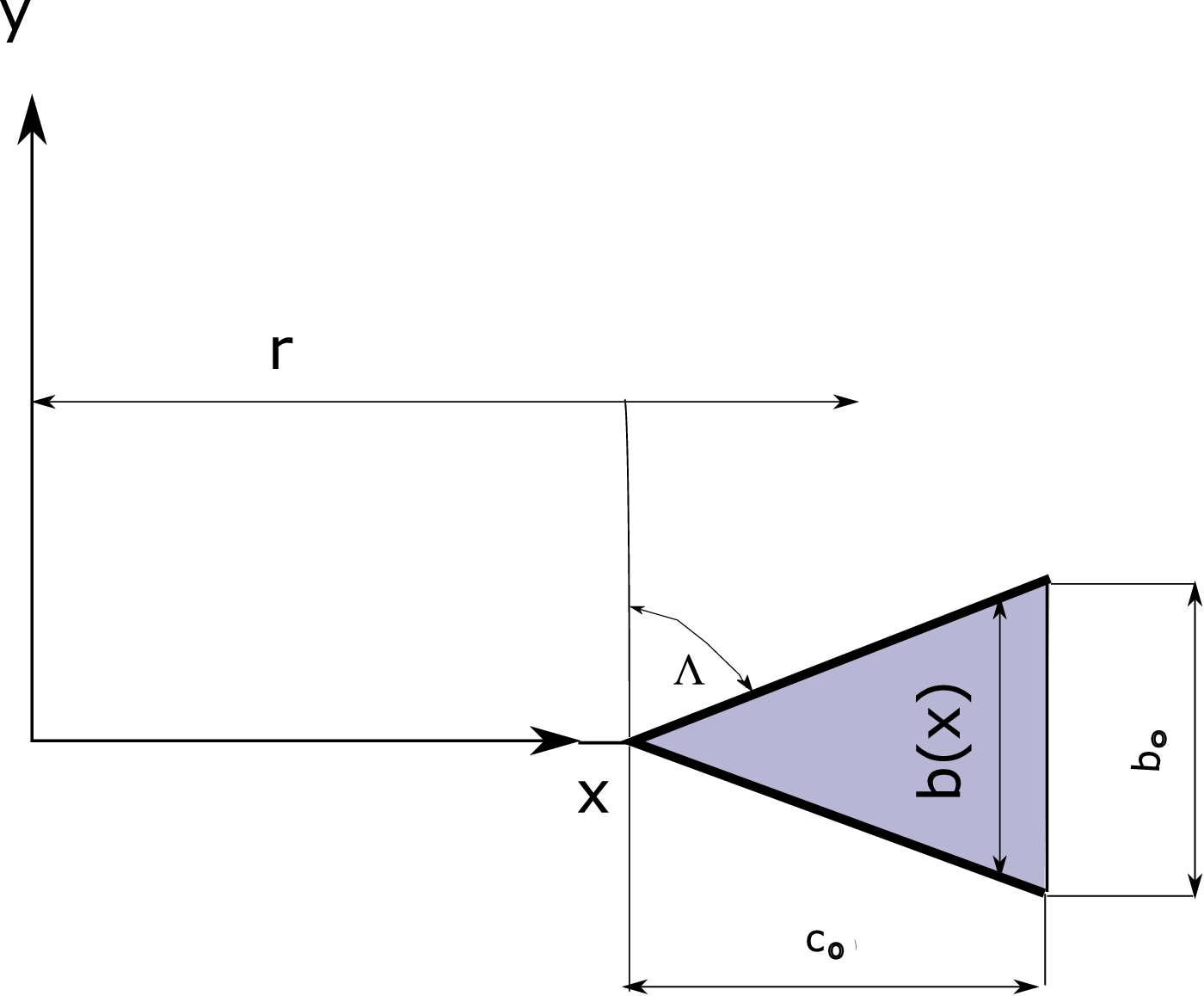}
\caption{Delta tail fin Schematic.  Figure taken from \citet{HW23}. }
\label{fig:3}
\end{figure}

\section{The nonlinear yaw response equation}
\label{sec:headings}
Figure \ref{fig:2} is a plan view of a generic tail fin yawing about the inertial axis (\textbf{x\textsubscript{o}},\textbf{y\textsubscript{o}}) fixed at \textbf{N} with a tail boom of length  $x_p$. For aeroelastic simulations of wind turbines, it is necessary to include the ``induced'' velocity, $V_{in}$, due to flow through the rotor, but $V_{in}=0$ in the TCs.  The ``elastic'' velocity $V_{el}$ arises  partly from any motion of the tower, which is absent in the TCs. $V_{el}$, however, includes  the rotational velocity of the tail fin which depends on the rate of change of $\gamma$, $\dot{\gamma}$, and is normal to the tail fin surface.  This velocity is the main difference between steady and unsteady yaw motion of tail fins and causes a difference between $\gamma$ and $\alpha$, the angle of attack in Fig.~\ref{fig:2}. 
All previous mathematical models of   fins or  vanes were derived  either from thin airfoil theory, following \citet{Wieringa1967}, or  unsteady slender body theory (USBT), following \citet{Wood2011}. Both use a small angle approximation to obtain a linear, second order ordinary differential equation in $\gamma$ by assuming that the vane rotational motion is slow compared to $U$, \citet{jonsson1988}. For the TCs, however, $\alpha$ and  $\gamma$  are related by 
\begin{equation}
   \tan \alpha= \arctan{\left(\tan \gamma + \frac{r \dot \gamma}{ U \cos \gamma}\right)}  \approx  \gamma+ \frac{r \dot \gamma}{ U }.
  \label{ang}
\end{equation}
The use of $\alpha$ to determine the forces  means that both theories mentioned above do not account properly for the suction force developed at the  sharp and swept leading edge of the fin. This suction force tends to rotate the resultant force in the direction normal to the surface but not normal to the flow direction as assumed in the two theories, \citet{robinson2013}. Hence the use of $\alpha$ for either resolving lift and drag in thin airfoil theory or for setting the boundary conditions in USBT is not correct and $\gamma$ is the  angle that gives the resultant normal force. The other fault of the linear theories is that both  neglect the nonlinear dependence of the normal force on $\gamma$ because they neglect the separation over all  thin leading edges at high sweep angle, $\Lambda$ as defined in Fig.~ \ref{fig:3}. The flow tends to separate into two concentrated spiral vortices even at low angle creating an additional component of induced velocity and nonlinear dependency on $\gamma$. Many theories have been developed to model the  effect of these vortices and supplement the basic linearized  theory on highly swept surfaces. All of them failed to some extent due to the assumptions used to make the solution manageable, mainly that the flow is conical such that it can be considered two dimensional. Then, \cite{Polhamus1966} formulated his leading edge suction analogy (LESA) which considered three dimensional flow and relaxed the assumption of small angle, \citet{Luckring2016}. The LESA was applied to model the tail fin response at high $|\gamma|$ by \citet{Ebert1995} and \citet{Singh12}. The reduced form of the high$-|\gamma|$ model is  based mainly on the LESA with some modifications to account for the complex high angle separation of swept thin edge surfaces. as  detailed in \citet{HW23}.   For analysing the TCs here, we limit $\gamma_0$ to $0 \le |\gamma_0| \le \pi/2$, as the limit angles admit the separate bounding linear solutions given in Subsection \ref{bound}.  The fin or vane is  released from rest at $\gamma_0$, giving  two initial conditions,  and the problem is to find the subsequent $\gamma(t)$, where $t$ is time.  A moment balance about $N$ gives
  \begin{equation}
   I \ddot{\gamma}+ M=0
  \label{1}
\end{equation}
where $ I$ is the  fin or vane moment of inertia about the yaw axis including the tail boom inertia. $M$ is the vane aerodynamic moment provided that friction can be ignored. The moment is taken from   Eq. (13.1) of \citet{HW23}: 
\begin{align}
I \ddot{\gamma}=&-0.5\rho A r \big[{x}_1^* K_{p}  \big(r \dot{\gamma}  
+ U  \sin \gamma   \big)U \cos{\gamma}  \nonumber \\ &+\big ({x}_2^* K_v +(1-{x}_3^*)C_{Dc}\big) \big(U\sin \gamma+r\dot{\gamma} \big)^2 \sign(U\sin \gamma+r\dot{\gamma}) \big]
\label{2}
\end{align}
where dots denote time derivatives and $\rho$ is the air density. The main assumption behind this ``reduced''  equation is that  $x_p \gg c_0$, where $c_0$ is the chord, defined in Fig.~\ref{fig:3}.  This restriction makes the yaw response of vanes and fins different from, say, a delta wing pitching about its apex for which $x_p=0$. The large literature on pitching bodies is, therefore, not of direct relevance to our analysis. The LESA, unfortunately, does not give the load distributions but only the integrated force.  \citet{HW23} used USBT to derive the yaw moment, which gave lengthy integrals, dependent on the chordwise direction, that are planform dependent. As an alternative, we assumed the  force obtained from LESA acts at  the moment arm $r$, taken to be the distance from the vane geometric centre to the yaw axis as shown in Fig.~\ref{fig:3}, and the  velocities are considered  uniform over the fin or vane.  These assumptions hold for most  fins and vanes known to the authors and was tested for some of them in \citet{kedr25}.  The first term in the brackets contains the potential flow coefficient $K_p$, and the second term expresses vortex flow, bursting and, ultimately, flow separation at very high yaw angle. Subsection \ref{bound} shows that these equations reduce to the linear form of \citet{Kris94} which involves only $K_p$, so the main nonlinearities are associated with the vortex flow coefficient, $K_v$ and $C_{Dc}$, the full separation drag coefficient.  Values of $K_p$ and $K_v$ will be considered and used throughout this study and it is very important to note that (a) most were obtained from lift and drag measurements or calculations for steady flow, and (b)
the unsteady values were obtained from system identification of wind tunnel results from model fins as described above.  ${x}_i^*$ for $i={1,2,3}$, are the separation functions. They express the contribution of each moment component as  $\gamma$  changes. 

For brevity, we do not reproduce the derivation of Eq.~(\ref{2}) from \citet{HW23}. The description in the last paragraph shows the major contribution of the LESA to the modelling of any planform; the reduction of the force acting on the these surfaces to  coefficients which are  planform and $\AR$ dependent extends the potential application of the model.  This was done, for example, by \citet{lamar1974} for a range of unusual planforms, and we note the success of the model for  fins of complicated planforms by \citet{kedr25}.  

\subsection {The yaw response as a second order nonlinear equation} \label{simpx}
Equation~(\ref{2}) is now recast in a form suitable for solution using averaging methods. First, the different flow components are assumed to be always active over the tail fin surface, so ${x}_1^*= {x}_2^* =1$. This implies that $C_{Dc}$ will behave identically to $K_v$, so we set ${x}_3^*=0$ and absorb the high$-|\gamma|$ drag into the vortex force coefficient. The reduced Eq.~(\ref{2}) now becomes the current ``minimal equation". In Section \ref{sec:coeff} it will be shown that these ${x}_{i}^*$ functions do affect the response and their modelling is important  for accurate prediction of the experimental response.  Our aim here, however, is to  understand how the model coefficients affect the response and this can be done using the minimal equation. Second, the moment of inertia is normalized to give the reduced moment of inertia defined by $I_*= 2I/( \rho A U^2 r)$ which has dimensions of s$^2$.  Equation (\ref{2}) becomes: 
\begin{multline}
  \ddot{\gamma} +\frac{1}{I_*}\left[\frac{r}{U}  {\cos \gamma}\dot{\gamma} +\frac{1}{2}{\sin 2\gamma}\right]K_p + \frac{1}{I_*} {\left[\sin {\gamma} +\frac{r \dot{\gamma}}{U}\right]}^2 \sign{\left[\sin {\gamma} +\frac{r \dot{\gamma}}{U}\right]}K_v=0.
  \label{3}
\end{multline}
Having two terms of different orders in the argument of the $\sign$ function makes it impossible to separate the terms containing $\gamma$ from those depending on $\dot {\gamma}$. 

\subsubsection {Polynomial expansion of the trigonometric functions} \label{poly}
 In order to have  Eq.~(\ref{3})  in standard form, the nonlinear trigonometric terms are transformed into polynomials. Using $\gamma^*= \gamma/\gamma_0$ as the normalized angle, the Chebyshev polynomials for sin(.) and cos(.) lead to
 \begin{equation}
  {\cos(\gamma_0 \gamma^*)}= J_0(\gamma_0)+2 \sum_{k=1}^{\infty} (-1)^k J_{2k}(\gamma_0) T_{2k}(\gamma^*)
  \label{6x}
\end{equation} and
 \begin{equation}
  {\sin(2\gamma_0 \gamma^*)}=2J_1(2\gamma_0)\gamma^*+ 2 \sum_{k=1}^{\infty} (-1)^k J_{2k+1}(2\gamma_0) T_{2k+1}(\gamma^*)
  \label{6x1}
\end{equation}
\citet{Snyder}. $J_n(.)$ is the Bessel function of the first kind of order $n$, and $T_k(.)$ is the  Chebyshev polynomial of order $k$.  (The names, definitions, and symbols for all the special functions used herein are taken from \citet{Olver2025} which will not be cited again.)

The expansion for $\sign$ term gives 
 \begin{flalign}
 \left[\sin( \gamma_0 \gamma^*) +r_u \gamma_0 \dot{\gamma}^*\right]^2&\sign{\left[\sin( \gamma_0 \gamma^*)+r_u \gamma_0 \dot{\gamma}^*\right]} \nonumber \\=&\frac{8}{3 \pi}\left[\sin( \gamma_0 \gamma^*) +r_u \gamma_0 \dot{\gamma}^*\right] -\frac{8}{\pi} \sum_{k=1}^{\infty}(-1)^{k}  \frac{T_{2k+1}(\sin( \gamma_0 \gamma^*) +r_u \gamma_0 \dot{\gamma}^*)}{(4k^2-1)(2k+3)}  \label{6x2}
\end{flalign}
where $r_u = r/U$, as derived in Appendix \ref{appA}.

\subsubsection {Series expansion of the yaw response}\label{seeq}

Substituting the trigonometric series expansions in  Eqs.~(\ref{6x}, \ref{6x1}, and \ref{6x2}) into Eq.~(\ref{3}) gives
\begin{eqnarray}
 \ddot{\gamma}^* &=& -\frac{K_p}{I_* \gamma_0}\left[ \sum_{k=0}^{\infty} (-1)^k J_{2k+1}(2\gamma_0) T_{2k+1}(\gamma^*)\right]-\frac{K_p r_u}{I_*}\left[ \sum_{k=0}^{\infty} \epsilon_k (-1)^k J_{2k}(\gamma_0) T_{2k}(\gamma^*)\right]{\dot{\gamma}^*}
 \nonumber\\  &&-\frac{K_v}{I_* \gamma_0}\left[\frac{8}{\pi} \sum_{k=0}^{\infty}(-1)^{k}  \frac{T_{2k+1}(\sin( \gamma_0 \gamma^*) +r_u \gamma_0 \dot{\gamma}^*)}{(4k^2-1)(2k+3)}\right] .
\label{eq43}
\end{eqnarray}
where $\epsilon_k=1$ for $k=0$, and $\epsilon_k=2$ for $k>0$.

To simplify the solution without comprising the accuracy, the terms in the last series summation in Eq.~(\ref{eq43}) of $\textit{O}({r}_u^3\dot{\gamma}^3)$ and higher are disregarded. This is shown in Appendix \ref{appB} to result in the simplified form of the response equation that is presented in the following subsection.

\subsection {Simplified response equation}\label{simp}
If the second term in the Chebyshev polynomial argument of the last term of Eq.~(\ref{eq43}) is small, then Appendix \ref{appB} derives the approximate equation correct to $\textit{O}({r}_u^3\dot{\gamma}^3)$. Thus,  Eq.~(\ref{3}) can be simplified to
\begin{multline}
  \ddot{\gamma} +\frac{1}{I_*}\left[r_u {\cos \gamma}\dot{\gamma} +\frac{1}{2}{\sin 2\gamma}\right]K_p + \frac{1}{I_*} \left[\sin {\gamma}|\sin {\gamma}| +2 r_u \dot{\gamma}|\sin {\gamma}|+r_u^2 \dot{\gamma}^2 \sign[{\gamma}]\right]K_v=0.
  \label{4}
\end{multline}
To check the accuracy of this approximation,  Eq.~(\ref{3}) and Eq.~(\ref{4}) were solved numerically. All the  numerical solutions presented herein were obtained using NDSolve in Mathematica without changing the default settings. A typical tail fin that is our TC1,  is taken from \citet{kedr23}.  It has a delta planform with $c_0= 0.27$ m and $b_0=0.078 $ m, which gave $\AR= 2b_0/c_0 = 0.58$, for which the force coefficients were $K_p= \pi \AR/2=0.91 $, and $K_v=\pi$.   $r=0.623$ m and $I$, the moment of inertia, was 0.047 kg m$^2$.  The comparisons were done at the maximum $\gamma_0=90^\circ$ and $U=17 $ m/s, giving $I_* = 0.042$ s$^2$. \citet{kedr23} showed that, for a fixed $\gamma_0$, the experimental response scaled with $U$, so that $r_u\dot{\gamma}$ will depend only on $\gamma_0$ and is maximized when $\gamma_0 = \pi/2$. 

\begin{figure}
\centering
\centerline{\includegraphics{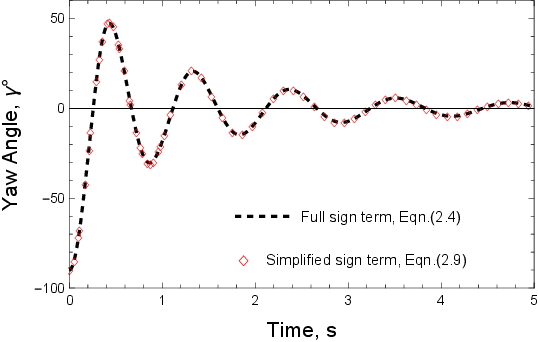}}
\caption{Yaw response of TC1 tail fin at $U = 17$ m/s. Solution using Eq.~(\ref{3}) is shown in dashed black line, and using Eq.~(\ref{4}) in red diamond symbols.}\label{fig:3x}
\bigbreak
\centerline{\includegraphics{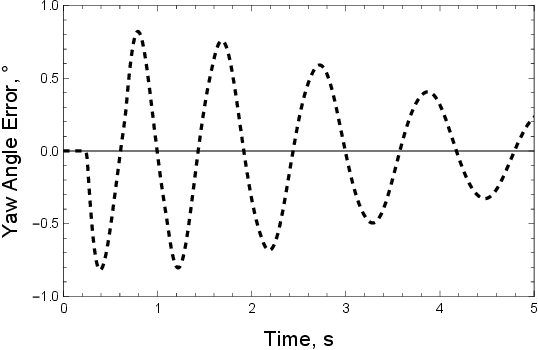}}
\caption{Difference between the 
   numerical solutions of Eqs.~(\ref{3}, \ref{4}) in figure~\ref{fig:3x}.}\label{fig:4x}
\end{figure}
The responses of the two equations are plotted in Fig.~\ref{fig:3x}. Figure~\ref{fig:4x} shows that the maximum error of the third order approximation in $r_u$ is around $\pm 0.81^\circ$, and the error decreases with time. The maximum value of $r_u\dot{\gamma}$ was $ 0.313$.  We conclude that Eq.~(\ref{4}) is sufficiently accurate to justify further analysis of it.
\subsection {The bounding linear solutions and the onset of nonlinearity} \label{bound}
 In the limit as $|\gamma| \to 0$ with $r_u\dot{\gamma}$ remaining small, the leading terms in Eq.~(\ref{4}) are
\begin{equation}
   \ddot{\gamma} + \frac{ r_u  K_p}{I_*} \dot{\gamma}+\frac{K_p}{I_*}  \gamma=0.
  \label{13b}
\end{equation}
Thus, the yaw response  is linearized and independent of $\gamma_0$.  As expected, Eq.~(\ref{13b}) is also independent of the nonlinear factor $K_v$. The solution of Eq.~(\ref{13b}) gives a damped oscillatory response with a natural frequency, $\omega_{0,\text{lin}}$, of
\begin{equation}
    \omega_{0,\text{lin}} = \sqrt{ K_p/I_*}
    \label{13c}
\end{equation}
and  damping ratio, $\zeta_{\text{lin}}$, of
\begin{equation}
   \zeta_{\text{lin}} = \frac{r_u}{2}\sqrt{\frac{K_p}{I_*}} \propto   r^{3/2} \sqrt{\frac{K_p A}{ I}}.
    \label{13cc}
\end{equation}and independent of $U$.  These are  Equations (40) and (41), respectively, for the frequency and damping for wind vanes in \citet{Kris94} where his $K$ is equivalent to $K_p$ here.

For completeness, we include the well-known solution to Eq.~(\ref{13b}):
\begin{equation}
    \gamma = \gamma_0 \exp(- \zeta_{\text{lin}}  \omega_{0,\text{lin}} t) \cos\left(\omega_{0,\text{lin}}\sqrt{1-\zeta_{\text{lin}}^2}  t\right).
    \label{13ccc}
\end{equation}
The largest $\gamma_0$ we consider is $\pi/2$.  With a change in variable to $\chi = \gamma_0 - \gamma$, Equation Eq.~(\ref{4})  becomes
\begin{equation}
   \ddot{\chi} - \frac{K_p}{I_*}\chi-\frac{K_v}{I_*}  =0
  \label{13d}
\end{equation}
when $r_u\dot{\gamma}$ and $\chi$ are small, \citet{kedr25}.  In terms of $\gamma$, the solution is

\begin{equation}
   \gamma= \gamma_0-\frac{K_v}{K_p} \left[\cosh \left(\sqrt{\frac{{K}_p}{I_*}}t\right)-1\right] \sim \gamma_0-\frac{K_v}{2I_*}t^2-\frac{K_p K_v}{24I_*^2}t^4- \textit{O}\left(\frac{t^6}{I_*^3}\right)
  \label{13e}
\end{equation}
and the response is clearly non-oscillatory which suggests that the nonlinearity quickly becomes important for high $\gamma_0$.

The lowest  order (nonlinear) terms that were omitted from the left hand side of Eq.~(\ref{13b}) are
\begin{equation}
   \frac{ K_v }{ I_*} \left( 2r_u \gamma \dot{\gamma} +  \  r_u^2|\dot{\gamma}|\dot{\gamma} +\gamma^2\right).
  \label{13f}
\end{equation}
Classical methods for the solution of nonlinear oscillations, e.g. \citet{Mickens1981}, require the nonlinear terms to be of the form $\epsilon F(\gamma, \dot{\gamma})$, where $F$ denotes functional dependence and  $\epsilon$ is a small perturbation parameter. Equation (\ref{4}) is not of this form as it has two parameters, neither of which is generally small. Further, the perturbation parameter in general will be a function of $\gamma_0$, but will be bounded by those in Eq.~(\ref{13f}) as $\sin \gamma$ is always smaller in magnitude than $\gamma$. 

As $\gamma_0 \rightarrow \pi/2$ the potential flow component diminishes and the $K_p$ term can be set to zero. The lowest order terms omitted from Eq.~(\ref{13d}) are 
\begin{equation}
   \frac{2 r_u K_v}{ I_* } \chi \dot{\chi} +  \frac{  r_u^2 K_v }{ I_*}|\dot{\chi}|\dot{\chi}
  \label{13g}
\end{equation}
based on the condition of $r_u \dot \gamma \ll 1$ which is  valid at the start of the response. This reduces the solution in Eq.~(\ref{13e}) to 
\begin{equation}
   \gamma \approx \gamma_0-\frac{ K_v }{2 I_*} t^2
  \label{chi}
\end{equation}
which  accurately reproduces the initial response  of a rectangular fin at high $|\gamma_0|$ in figure 3 of \citet{kedr25}. As expected, the  high$|\gamma_0|$ response is dominated initially by $K_v$.

\subsection {The response equation in nondimensional form}
\label{standard}

  There are two possible parameters which would nondimensionalize the time, $t$; $\sqrt{I_*}$ or $r_u$.     We chose to transform $t$ to $t_* = t/\sqrt{I_*}$ as suggested by Eq.~(\ref{chi}). This makes the coefficient of $d^2 \gamma/d t_*^2$ equal to unity. As stated previously,  $\gamma$ is nondimensionalised by $\gamma_0$ to give $\gamma^*= \gamma/\gamma_0$. Equation (\ref{4}) becomes
\begin{align}
  \ddot{\gamma}^* &+ \frac{{\beta_1}}{2}{\sin (2 \gamma_0 \gamma^*)+{\beta_2} {\cos( \gamma_0 \gamma^*)}{\dot{\gamma}^*}} \nonumber \\&+
  {\beta}{\beta_1} {\left[\left(\sin( \gamma_0 \gamma^*)^2  +\frac{2\beta_2 \dot{\gamma}^*}{\beta_1}\right) \sign{\left[\sin( \gamma_0 \gamma^*)\right]} +\left(\frac{\beta_2 \dot{\gamma}^*}{\beta_1}\right)^2 \sign{\left[ \gamma_0 \gamma^*\right]}\right]} =0.
  \label{5}
\end{align}
where ${\beta}= K_v/K_p$, ${\beta_1}= K_p/\gamma_0$, and ${\beta_2}= K_p r_u/\sqrt{I_*}$ are the  parameters that will be investigated in the Section 6. Because these parameters appear throughout the remaining text, they are given for ease of reference in Table \ref{tab1}. The initial conditions are $\gamma^{*}{(t=0)}= 1$ and $\dot{\gamma}^{*}{(t=0)}= 0$.
To standardize the response equation,  damping and natural frequency terms with coefficients $2 \sigma$ and ${\nu}^2$,  respectively, are added to the left side of Eq.~(\ref{5}) and equivalent terms with coefficients $\epsilon \sigma^*$, and $\epsilon {\omega^*}^2$ are added to the right side to give
\begin{equation}
  \ddot{\gamma}^* + 2 \sigma \dot{\gamma}^*+  {\nu}^2 \gamma^*= \epsilon F( \gamma^*,\dot{\gamma}^*).
  \label{6}
\end{equation}
where $\sigma$, $\nu$, are  the equivalent linear damping and frequency to be found during the solution process.  The nonlinear right hand side $\epsilon F( \gamma^*,\dot{\gamma}^*)$ is a function of the  perturbation parameter, $\epsilon$ which will be determined in Section \ref{sec:epsilon} and  shown to be a  linear function of $\beta_2$.  Thus, the the last term in the last bracket of Eq.~(\ref{5}) is  of second order and can be neglected. $\epsilon F(\gamma^*,\dot{\gamma}^*)$ is given by
\begin{align}
  \epsilon F( \gamma^*,\dot{\gamma}^*)=& -\frac{{\beta_1}}{2}{\sin (2 \gamma_0 \gamma^*)-{\beta_2} {\cos( \gamma_0 \gamma^*)}{\dot{\gamma}^*}} \nonumber\\&-
  {\beta}{\beta_1} {\left(\sin( \gamma_0 \gamma^*)  +\frac{2\beta_2 \dot{\gamma}^*}{\beta_1} \right)} |\sin( \gamma_0 {\gamma}^*)| + \sigma^* \dot{\gamma}^*+{{\omega}^*}^2 {\gamma}^*.
  \label{7}
\end{align}
\begin{ruledtabular}
\begin{table}
\caption{\label{tab1}%
The main parameters of the minimal equation. $I_*$ and $r_u$ have dimensions s$^2$ and s respectively.  The remaining parameters have no dimensions.
}
 \begin{center}
\def~{\hphantom{0}}
  \begin{tabular}{lccccccc}
   $I_*= 2I/( \rho A U^2 r)$&$r_u=r/U$&  ${\beta}= K_v/K_p$& ${\beta_1}= K_p/\gamma_0$&${\beta_2}= K_p r_u/\sqrt{I_*}$&$t_*=t/I_*$&$\gamma^* = \gamma/\gamma_0$ \\
  \end{tabular}
  \end{center}
\end{table}
\end{ruledtabular}
\section{Solution of the yaw response equation}
\label{sec:solution}

Perturbation methods are powerful techniques for obtaining approximate closed form solutions to nonlinear second order differential equations of the forms studied here. Three basic perturbation methods are most commonly used,  namely: the Lindstedt-Poincare method, the averaging methods, and the the method of multiple scales, \citet{Mickens1981}.   Of these, the KBM method, due to \citet{Krylov47} and \citet{Bog61}, is the most suitable for the present analysis. 
A major advantage of  averaging  is that it transforms the complicated second nonlinear equation into two separable first-order equations which are straightforward to solve. In contrast, the multiple scale method leads to partial differential equations which can get complicated for complex nonlinear equations. A further advantage of the averaging method is that the solution is not determined by the need to get higher orders of the series expansion, but by ensuring the asymptotic behaviour as the perturbation parameter $\epsilon \to 0$ for each term of the series. This makes the determination of the higher order terms of the series independent of the higher harmonics in contrast to the Lindstedt-Poincare method. 

In this section, we describe the application of the KBM method of averaging for the undamped system, as detailed in the Appendix \ref{appC}, and as extended by \citet{Mendelson1970} for finite damping.

 Following the KBM method, we seek a solution that reduces to its linear equivalent in the limit $ \epsilon F( \gamma^*,\dot{\gamma}^*) \to 0$. The solution has the form 
\begin{equation}
 {\gamma}^*=  {\gamma}^*(a,\psi)={\gamma}^{*}_{0}+ \epsilon {\gamma}^{*}_{1}+ {\epsilon}^2 {\gamma}^{*}_{2}
  \label{15}
\end{equation}
where $a$ is the amplitude and $\psi$ is the phase angle of period $2\pi$. Note that the subscript ``0'' indicates the linear solution for  $\gamma^*$ in the limit $\epsilon \rightarrow 0$, and not the initial condition on $\gamma$.  The first order approximation to Equation Eq.~(\ref{15}) is 
\begin{equation}
 {\gamma}^*={\gamma}^{*}_{0}=  a \cos {\psi}.
  \label{16}
\end{equation}
For the linear solution of Eq.~(\ref{13ccc}), $a= \exp(- \zeta_{\text{lin}}  \omega_{0,\text{lin}} t)$ and $\psi =\omega_{0,\text{lin}}(1-\zeta_{\text{lin}}^2)^{1/2}  t$.  The frequency, $\omega=d\psi/dt=\omega_{0,\text{lin}}(1-\zeta_{\text{lin}}^2)^{1/2}.$ The relation $\omega=d\psi/dt$ holds for all linear and nonlinear systems. To first order in the perturbation parameter $\epsilon$,  $a$ and $\psi$ are given by first order differential equations:
\begin{equation}
 \frac{da}{dt}=-\sigma a +\epsilon \zeta_1(a).
   \label{17}
\end{equation}
and
\begin{equation}
 \frac{d \psi}{dt}=\omega_0 +\epsilon \omega_1(a)
   \label{18}
\end{equation}
such that $\omega_0= \sqrt{\nu^2-\sigma^2}$. The following assumption is used in the KBM method, \citet{Mickens1981}: 
\begin{equation}
 \zeta_1(a) \cos {\psi} - a \omega_1(a) \sin {\psi} =0  .
   \label{n1}
\end{equation}
This restriction is necessary for simplifying  $\dot{\gamma}^*$ to
\begin{equation}
\dot{\gamma}^* = -( \sigma \cos {\psi}+ \omega_0 \sin {\psi}) a .
   \label{n2}
\end{equation}
The consequences of this assumption and its impact on the accuracy of the solution will be assessed in Section \ref{sec:ARC}.

The unknown functions $\zeta_1(a)$ and $\omega_1(a)$ are determined from the condition that Eq.~(\ref{15}) satisfies Eq.~(\ref{6}) for each order of $\epsilon$. The leading terms in  Eq.~(\ref{17}) and  Eq.~(\ref{18}) are chosen such that  the linear solution is recovered as $\epsilon  F( \gamma^*,\dot{\gamma}^*) \to 0$.  The resulting systems of second order equations in $\zeta(a)$ and $\omega(a)$ can be solved by expressing these variables as a power series in $a$: 
\begin{equation}
 \zeta_1(a)=\sum_{k=2}^{2m+1} {A_k^{(1)}} a^k
   \label{19}
\end{equation}
and
\begin{equation}
 \omega_1(a)=\sum_{k=2}^{2m+1} {B_k^{(1)}}a^{k-1}
   \label{20}
\end{equation}
where $m$ depends on the degree of the truncated terms of the series. For  most systems, the first order solution given by  
 Eq.~(\ref{16}) and in Eqs.~(\ref{17},\ref{18}) are sufficient to represent the approximate solution of Eq.~(\ref{15}), \citet{Bojadziev1980}.  We also confirmed that the higher order terms were negligible. This occurs generally when the nonlinear function in  Eq.~(\ref{6}) contains terms of odd degrees which is indeed the case for Eq.~(\ref{5}).

The coefficients $A_k$ and $B_k$ are obtained from the requirement that the secular terms vanish for any value of $\sigma$, see Appendix \ref{appC}. Upon substituting  Eqs.~(\ref{16}, \ref{19}, and \ref{20}) into  Eq.~(\ref{5}) and expanding $\epsilon  F( \gamma^*,\dot{\gamma}^*) $ in  Eq.~(\ref{6}) as a Fourier series of $\gamma^*$ to give
\begin{equation}
 2 \omega_0 \omega_1 a+ \sigma a \frac{d\zeta_1}{da}- \sigma \zeta_1 = -F_1(a)
   \label{21}
\end{equation}
and
\begin{equation}
 2 \omega_0 \zeta_1- \sigma a^2 \frac{d\omega_1}{da} = -G_1(a)
   \label{22}
\end{equation}
where $F_1(a)$ and $G_1(a)$ are, respectively, the even and odd first coefficients of the Fourier series expansion of $\epsilon F( \gamma^*,\dot{\gamma}^*)$ given as
\begin{equation}
 F(a \cos {\psi},-\omega_0 a \sin {\psi}-\sigma a \cos {\psi}) = \frac{F_0(a)}{2}+\sum_{n=1}^{\infty} \big[F_n(a) \cos {n \psi}+G_n(a) \sin {n \psi}\big]. 
   \label{23}
\end{equation}
The coefficients $F_n(a)$, $G_n(a)$ can be expanded in a series in the amplitude $a$ as:
\begin{align}
&F_n(a) = \sum_{k=1}^{2m+1}  F_k^{(n)} a^k, \hspace{2mm} \text{and}\\
 &G_n(a) = \sum_{k=1}^{2m+1}  G_k^{(n)} a^k.
   \label{24}
\end{align}
Substituting Eqs.~(\ref{19} and \ref{20}) into Eqs.~(\ref{21} and \ref{22}) and retaining the first harmonics ($n=1$) of Eq.~(\ref{23}) results in the following equations for the coefficients of $\zeta_1$ and $\omega_1$:
\begin{equation}
 {A_k^{(1)}}= -\frac{(k-1)\sigma F_k^{(1)}+2 \omega_0 {G_k^{(1)}}}{4{\omega_0}^2+(k-1)^2\sigma^2}
   \label{25}
\end{equation}
and
\begin{equation}
 {B_k^{(1)}}=  -\frac{2 \omega_0 F_k^{(1)}-(k-1)\sigma{G_k^{(1)}}}{4{\omega_0}^2+(k-1)^2\sigma^2} .
 \label{26}
 \end{equation}
Substituting Eqs.~(\ref{25} and \ref{26}) into  (\ref{19} and \ref{20}) allows the solution of Eqs.~(\ref{17} and \ref{18}).

\subsection {Application of the averaging method to the yaw response equation}\label{subsec:solution}

In order to solve  Eqs.~(\ref{17} and \ref{18}) analytically, the value of $m=2$ is used. This makes the RHS of Eq.~(\ref{17}) a quintic polynomial in $a$. Equations (\ref{17} and \ref{18}) are 
\begin{equation}
 \frac{da}{dt}=-\sigma a+ \epsilon {A}_3^{(1)} a^3 + \epsilon {A}_5^{(1)} a^5
   \label{27}
\end{equation}
and
\begin{equation}
 \frac{d \psi}{dt}={\omega}_0 +\epsilon B_3^{(1)} a^2 + \epsilon B_5^{(1)} a^4.
   \label{28}
\end{equation}

The Fourier coefficients for $n=1$, $F_i^{(1)}$ and $ G_i^{(1)}$ of $ F( \gamma^*,\dot{\gamma}^*)$  as expressed by  Eq.~(\ref{23}) were calculated as
\begin{equation}
F_1(a)= \frac{1}{\pi}\int_0^{2\pi} {F(a \cos {\psi},-\omega_0 a \sin {\psi}-\sigma a \cos {\psi}) \cos {\psi} d\psi}=F_1^{(1)} a+ F_3^{(1)} a^3+F_5^{(1)} a^5
\label{f1}
\end{equation}
and
\begin{equation}
G_1(a)=\frac{1}{\pi}\int_0^{2\pi} {G(a \cos {\psi},-\omega_0 a \sin {\psi}-\sigma a \cos {\psi}) \sin {\psi} d\psi}=G_1^{(1)} a+ G_3^{(1)} a^3+G_5^{(1)} a^5
\label{f2}
\end{equation}
where all the even coefficients vanish. The coefficients $F_1(a)$ and $G_1(a)$ are obtained by integrating the nonlinear function from Eq.~(\ref{7}) using Mathematica as
\begin{eqnarray}
F_1(a)=&&-\beta _1 J_1\left(2 \gamma_0a\right)+ \frac{2 {\beta }_2 \sin \phi }{{\gamma}_0} \left(J_1\left(\gamma_0  a\right)- a {\gamma}_0J_2\left(\gamma_0  a\right)\right)\nonumber\\ &&-{\beta}{\beta} _1\sign\left(\gamma _0\right)\left[ \pmb{H}_1\left(2 a \gamma _0\right)+\frac{4 \beta _2 \sin \phi \left(a \gamma _0 \pmb{H}_0\left(a \gamma _0\right)-\pmb{H}_1\left(a \gamma _0\right)\right)}{\beta _1 \gamma _0}\right]\nonumber\\ && -(\sigma ^* \sin \phi- {\omega^*}^2)a
\label{32}
\end{eqnarray}
and
\begin{flalign}
 G_1(a)= \frac{2 {\beta} _2 \cos\phi }{{\gamma}_0}  J_1\left(\gamma_0a\right)+{\beta}{\beta} _2\sign\left(\gamma _0\right)\left[\frac{4     \pmb{H}_1\left(a \gamma _0\right)\cos{\phi}}{\gamma _0} \right]- \sigma ^* a \cos \phi
   \label{33}
\end{flalign}
where $\pmb{H}_0(.)$ and $\pmb{H}_1(.)$ are Struve functions. The terms in the first and the second lines result from integrating the respective terms in Eq.~(\ref{7}). In deriving the equations, the yaw rate from Eq.~(\ref{n2}) is set to the form $\dot{\gamma}^* = -\sin(\psi+\phi) a $, such that $\sin\phi= \sigma$, and $\cos\phi= \omega_0$. As $\sin \phi \approx \textit{O}(\beta_2)$, the second term in the brackets of Eq.~(\ref{32}) is of the order $\textit{O}( {\beta}_2^2)$ and is negligible as found before. This results in a simpler expressions of the Fourier coefficients as
\begin{eqnarray}
&&F_1(a)=-\beta _1\left[ J_1\left(2 \gamma_0a\right)+{\beta \sign\left(\gamma _0\right) \pmb{H}_1\left(2 a \gamma _0\right) }\right]+ {\omega^*}^2a
\label{32s}
\end{eqnarray}
and
\begin{eqnarray}
&&G_1(a)=\frac{2 \beta_2 \cos\phi}{\gamma_0} \left[  J_1\left(\gamma_0a\right)+2{\beta}\sign\left(\gamma _0\right)\pmb{H}_1\left(a \gamma _0\right) \right]-\sigma ^* a \cos\phi
   \label{33s}
\end{eqnarray}
\subsection {Analytical representation of the response equation}
 
KBM requires the Fourier coefficients of the nonlinear function to be in a polynomial form. The functions $F_1(a)$ and $G_1(a)$ can be approximated as polynomials in the amplitude allowing analytical solution using approximate methods, \citet{Beléndez}. As the argument spans  $a \in [0,1]$, we used Chebyshev  polynomials of the first kind, $T_{2n+1}(a)$.  The approximation is most accurate over the interval $\pi/4 < |\gamma| < \pi/2$, \citet{Teron}, which makes it immediately suitable for  high $|\gamma|$. 

In order to get the $F_1$ and $G_1$ functions in a polynomial  of $a$,  the Bessel functions are expanded in a Chebyshev series of $a$ as 
\begin{eqnarray}
J_{2 k+1}\left(\gamma _0 a\right)= 2 \sum _{n=0}^m  J_{k+n+1}\left(\gamma _0/2\right) J_{k-n}\left(\gamma _0/2\right) T_{2 n+1}(a).
   \label{eqnj}
\end{eqnarray}
The Struve function expansion in terms of Chebyshev series is derived from the  Schl{\"o}milch series of non alternating sign stated in \citet[Section 3.B]{Actor1987} as 
\begin{eqnarray}
\pmb{H}_1\left(a \gamma _0\right)=\frac{2 \gamma _0^2}{3 \pi }\sum _{n=0}^m \frac{ \, _2F_3\left(2,1;5/2,n+5/2,3/2-n;-\gamma _0^2/4\right)}{\Gamma \left(3/2-n\right) \Gamma \left(n+5/2\right)}T_{2 n+1}(a)
   \label{eqnH}
\end{eqnarray}
where $_2F_3$ is the generalized hypergeometric function, and $\Gamma(.)$ is the Gamma function. To express the Fourier coefficients as a polynomial in $a$, the odd Chebyshev polynomial $T_{2n+1}$ is expressed in terms of the hypergeometric function $_2F_1$,  \citet[Eq. (9), Sec.8.5.1]{Luke1969}:
\begin{equation}
T_{2 n+1}(a)= (-1)^n (2n+1) a \, _2F_1\left(-n,n+1;3/2;a^2\right).
\label{T1}
\end{equation}
$_2F_1$ has a terminating series if either the first or the second arguments are non positive integers which gives
\begin{equation}
T_{2 n+1}(a)= (-1)^n (2n+1) a \sum _{k=0}^n \frac{(-1)^k  \Gamma (k+n+1)}{\Gamma (2 k+2) \Gamma (-k+n+1)} (2 a)^{2 k}.
\label{T2}
\end{equation}
Similarly to the Struve function in Eq.~(\ref{eqnH}), the Bessel function in Eq.~(\ref{eqnj}) is expressed in terms of $_2F_3$, using the identity of Bessel functions products from  \citet[Eq. (39), Sec.6.2.7]{Luke1969} and replacing them in Eq.~(\ref{eqnj}) to give
\begin{eqnarray}
{J}_1\left(a \gamma _0\right)=\left(\frac{\gamma _0}{2}\right)\sum _{n=0}^m \frac{ (-1)^n \, _2F_3\left(1+n,3/2+n;n+2,n+1,2n+2;-\gamma_0^2/4\right)}{\Gamma \left(n+2\right) \Gamma \left(n+1\right)} \left(\frac{\gamma _0}{4}\right)^{2n}T_{2 n+1}(a).\nonumber \\
   \label{eqnj2}
\end{eqnarray}
To the authors' knowledge, the series representations of the $\pmb{H}$ and $J$ functions of the first kind in Eqs.~(\ref{eqnH} and \ref{eqnj2}) respectively are derived herein for the first time.  They are valid only for $a \ge0$. Substituting Eqs.~(\ref{eqnH}, \ref{T2}, and \ref{eqnj2}) in (\ref{32s} and \ref{33s}) gives $F_1(a)$ and $G_1(a)$ as
\begin{eqnarray}
F_1(a)&=&-a \beta_1 \sum _{n=0}^m (2n+1) \Bigg[  2 \Gamma \left(2n+2\right)  \, {_2}\tilde{ F}_3\left(1+n,3/2+n;n+2,n+1,2n+2;- \gamma _0^2\right)\nonumber \\&&\times\left(\frac{\gamma _0}{2}\right)^{2n+1}+ \frac{2 {\beta}|\gamma _0|\gamma _0}{ \sqrt{\pi} } (-1)^n \,{_2}\tilde{ F}_3\left(2,1;5/2,n+5/2,3/2-n;- \gamma _0^2\right)\Bigg]\nonumber\\ &&\times\sum _{k=0}^n \frac{(-1)^k  \Gamma (k+n+1)}{\Gamma (2 k+2) \Gamma (-k+n+1)} (2 a)^{2 k}+ {\omega^*}^2a
\label{F1}
\end{eqnarray}
and
\begin{align}
G_1(a)=& a \beta_2 \cos\phi \sum _{n=0}^m (2n+1) \Bigg[\Gamma \left(2n+2\right) \nonumber\\ &\times {_2}\tilde{ F}_3\left(1+n,3/2+n;n+2,n+1,2n+2;-\gamma_0^2/4\right)\left(\frac{\gamma _0}{4}\right)^{2n} +\frac{2 {\beta}|\gamma _0|}{ \sqrt{\pi} } (-1)^n \, \nonumber\\ &\times{_2}\tilde{ F}_3\left(2,1;5/2,n+5/2,3/2-n;-\gamma_0^2/4\right)\Bigg] \sum _{k=0}^n \frac{(-1)^k  \Gamma (k+n+1)}{\Gamma (2 k+2) \Gamma (-k+n+1)} (2 a)^{2 k}\nonumber\\ &-\sigma ^* a\cos\phi
   \label{G1}
\end{align}
where ${_2}\tilde{ F}_3\left(a_1,a_2;b_1,b_2,b_3;z\right)= {_2}F_3\left(a_1,a_2;b_1,b_2,b_3;z\right)/\Gamma(b_1)/\Gamma(b_2)/\Gamma(b_3)$ is the ``regularized'' $_2F_3$, \citet{Straton2024}. These formulations of the Fourier coefficients allow their straightforward expression in ascending order of $a$. To approximate $F_1$ and $G_1$ as fifth order polynomials in $a$, $m=2$ is set in Eqs.~(\ref{F1} and \ref{G1}). Also, the power series coefficients of $a$, $ F_{2k+1}^{(1)}$ and $ G_{2k+1}^{(1)}$ in Eq.~(\ref{24}) are easily obtained by setting $k = 0,1,2$ in Eqs.~(\ref{F1}, \ref{G1}). When $k=0$, the coefficients of $a$ are  
\begin{eqnarray}
 F_1^{(1)} &=& - \beta_1 \sum _{n=0}^m (2n+1) \Bigg[ 2 \Gamma \left(2n+2\right)  \, {_2}\tilde{ F}_3\left(1+n,3/2+n;n+2,n+1,2n+2;- \gamma _0^2\right)\nonumber\\ &&\times\left(\frac{\gamma _0}{2}\right)^{2n+1}+ \frac{2 {\beta}|\gamma _0|\gamma _0}{ \sqrt{\pi} } (-1)^n \,{_2}\tilde{ F}_3\left(2,1;5/2,n+5/2,3/2-n;- \gamma _0^2\right)\Bigg]+ {\omega^*}^2,\hspace{2mm}\text{and}\nonumber\\
G_1^{(1)}&=& \beta_2 \cos\phi \sum _{n=0}^m (2n+1)\Bigg[\Gamma \left(2n+2\right)  \, {_2}\tilde{ F}_3\left(1+n,3/2+n;n+2,n+1,2n+2;-\gamma_0^2/4\right)\nonumber \\&&\times\left(\frac{\gamma _0}{4}\right)^{2n}+\frac{2 {\beta}|\gamma _0|}{ \sqrt{\pi} } (-1)^n \, {_2}\tilde{ F}_3\left(2,1;5/2,n+5/2,3/2-n;-\gamma_0^2/4\right)\Bigg]\nonumber \\&&-\sigma^*\cos\phi
 \label{eqn2}
\end{eqnarray}
Setting $F_1^{(1)}=0$ and $G_1^{(1)}=0$, the equivalent linear damping and frequency in Eq.~(\ref{6}) are obtained as 
\begin{align}
\sigma&=\frac{\sigma ^*}{2}\nonumber\\&=\frac{\beta _2}{2}\sum _{n=0}^m (2n+1) \Bigg[ \, \Gamma \left(2n+2\right)  \, {_2}\tilde{ F}_3\left(1+n,3/2+n;n+2,n+1,2n+2;-\gamma_0^2/4\right)\left(\frac{\gamma _0}{4}\right)^{2n}\nonumber\\ &~~~~~+\frac{2 {\beta}|\gamma _0|}{ \sqrt{\pi} } (-1)^n \, {_2}\tilde{ F}_3\left(2,1;5/2,n+5/2,3/2-n;-\gamma_0^2/4\right) \Bigg]
   \label{eqn4}
\end{align}and
\begin{align}
{\nu}^2&={\omega^*}^2\nonumber\\&= \beta_1 \sum _{n=0}^m (2n+1) \Bigg[ 2 \Gamma \left(2n+2\right)  \, {_2}\tilde{ F}_3\left(1+n,3/2+n;n+2,n+1,2n+2;- \gamma _0^2\right)\left(\frac{\gamma _0}{2}\right)^{2n+1}\nonumber\\ &~~~~~+ \frac{2 {\beta}|\gamma _0|\gamma _0}{ \sqrt{\pi} } (-1)^n \,{_2}\tilde{ F}_3\left(2,1;5/2,n+5/2,3/2-n;- \gamma _0^2\right)\Bigg].
   \label{eqn5}
\end{align}
The remaining coefficients are 
\begin{eqnarray}
F_3^{(1)}&=& \frac{2 \beta_1}{3}  \sum _{n=1}^m n(n+1)(2n+1) \Bigg[  2 \Gamma \left(2n+2\right)  \, {_2}\tilde{ F}_3\left(1+n,3/2+n;n+2,n+1,2n+2;- \gamma _0^2\right)\nonumber\\ && \times\left(\frac{\gamma _0}{2}\right)^{2n+1}+\frac{2 {\beta}|\gamma _0|\gamma _0}{ \sqrt{\pi} } (-1)^n \,{_2}\tilde{ F}_3\left(2,1;5/2,n+5/2,3/2-n;- \gamma _0^2\right)\Bigg],\nonumber\\ 
F_5^{(1)}&=& 16 \beta_1 \left[\frac{64 \beta  \gamma _0 \left| \gamma _0\right|  \, _2F_3\left(1,2;-1/2,5/2,9/2;-\gamma _0^2\right)}{315 \pi ^2}-\frac{\gamma _0^5}{192} {_1}{ F}_2\left(7/2;4,6;-\gamma _0^2\right)\right],\nonumber\\ 
 G_3^{(1)}&=&  \frac{-2 \beta_2 \cos\phi }{3} \sum _{n=1}^m n(n+1)(2n+1) \Bigg[   \, \Gamma \left(2n+2\right)  \\ &&\times {_2}\tilde{ F}_3\left(1+n,3/2+n;n+2,n+1,2n+2;-\gamma_0^2/4\right)\nonumber\left(\frac{\gamma _0}{4}\right)^{2n}\nonumber \\&&+ \frac{2 {\beta}|\gamma _0|}{ \sqrt{\pi} } (-1)^n \, {_2}\tilde{ F}_3\left(2,1;5/2,n+5/2,3/2-n;-\gamma_0^2/4\right) \Bigg],\hspace{2mm}\text{and} \nonumber\\
 G_5^{(1)}&=& 16 \beta _2 \cos\phi \left[\frac{\gamma _0^4 \, _1F_2\left(7/2;4,6;-\frac{\gamma _0^2}{4}\right)}{3072}-\frac{64 \beta  \left| \gamma _0\right|  \, _2F_3\left(1,2;-1/2,5/2,9/2;-\frac{\gamma _0^2}{4}\right)}{315 \pi ^2}\right]. \nonumber
  \label{eqn6}
\end{eqnarray}
\section{Nonlinear perturbation parameter and error analysis}
\label{sec:epsilon}
In this section, an expression of the perturbation parameter $\epsilon$ is developed and an estimation of the error resulting from truncating the series of the coefficients Eq.~(\ref{F1}) and (\ref{G1}) are stated.

\begin{figure}
  \centering
     \begin{subfigure}[b]
     {0.495\textwidth}
         \centering
       \includegraphics[width=\textwidth]{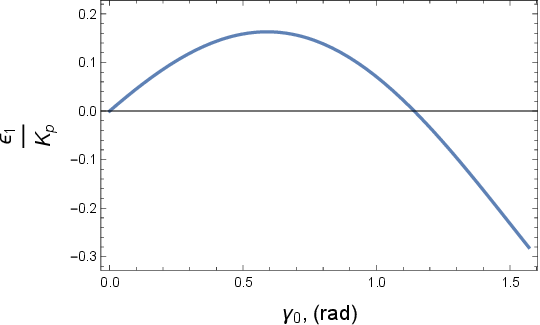}
       \caption{}
     \end{subfigure}
      \begin{subfigure}[b]
      {0.495\textwidth}
         \centering
       \includegraphics[width=\textwidth]{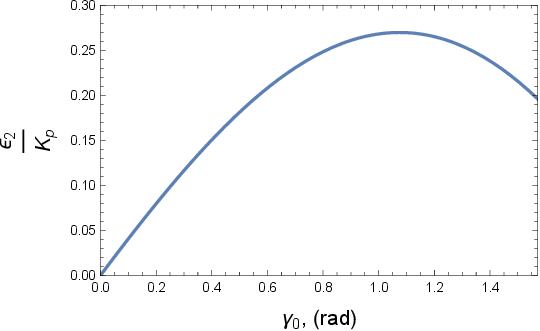}
       \caption{}
     \end{subfigure}
     \caption{Variation of the perturbation parameters change with $\gamma_0$ for TC1 described at the beginning of subsection \ref{simp}. a) $\epsilon_1/K_p$, b) $\epsilon_2/K_p$.} 
\label{fig:e1}
\end{figure}

The terms in the nonlinear function $\epsilon F( \gamma^*,\dot{\gamma}^*)$ from Eq.~(\ref{7}) are expressed as Chebyshev series to separate the $\gamma^*$ terms using the expansions in Eqs.~(\ref{6x} and \ref{6x1}).  We use the following series expansions for the sin terms: 
\begin{equation}
  {|\sin(\gamma_0 \gamma^*)|}=- \left[E_{0}(\gamma_0)+2\sum_{k=1}^{\infty}(-1)^k  E_{2k}(\gamma_0)  T_{2k}( \gamma^*)\right] 
  \label{nnl1}
\end{equation}
\begin{eqnarray} |\sin(\gamma_0\gamma^*)|\sin(\gamma_0\gamma^*)&=& \left(\frac{2}{\pi}-E_{1}(2\gamma_0)\right)\gamma^*\nonumber\\ &&+\sum_{k=1}^{\infty}(-1)^k  \left( \frac{ 2}{\pi(2k+1)}-E_{2k+1}(2\gamma_0) \right)  T_{2k+1}(\gamma^*) 
  \label{nnl2}
\end{eqnarray}
where $E_k(.)$ are Weber functions of order $k$ as derived in Appendix \ref{appD}. The series expansion for $T_{2k+1}$ is given by Eq.~(\ref{T2}) and for $T_{2k}$ by
\begin{equation}
T_{2 k}({\gamma}^*)= (-1)^k+(-1)^k k \sum _{n=1}^k \frac{(-4)^n  \Gamma (k+n)}{\Gamma (2 n+1) \Gamma (-n+k+1)} ( {\gamma}^*)^{2 n}
\label{nnl3}
\end{equation}
are substituted above to separate $\epsilon F( \gamma^*,\dot{\gamma}^*)$ into nonlinear, $\epsilon F( \gamma^*,\dot{\gamma}^*)_{nl}$, and linear terms.
\begin{eqnarray}
 \epsilon F( \gamma^*,\dot{\gamma}^*)_{nl}&=&\beta_1 \sum _{k=1}^\infty  (-1)^k \left[J_{2k+1}(2\gamma_0) +\beta \left( \frac{ 2}{\pi(2k+1)}-E_{2k+1}(2\gamma_0) \right)\right] S_1 \nonumber\\&& +2\beta_2 \sum _{k=1}^\infty (-1)^k \left[J_{2k}(\gamma_0) -2\beta E_{2k}(\gamma_0)\right]S_2.
  \label{nnl5}
\end{eqnarray}
where $S_1$ and $S_2$, the series summation of $\gamma^*$ from Eqs.~(\ref{T2} and \ref{nnl3}), are given respectively as  
\begin{eqnarray}
S_1 &=& (-1)^k (2k+1) \sum _{n=1}^k \frac{(-4)^n  \Gamma (k+n+1)}{\Gamma (2 n+2) \Gamma (-n+k+1)} ( {\gamma}^*)^{2 n+1}\nonumber\\
S_2 &=& (-1)^k k \sum _{n=1}^k \frac{(-4)^n  \Gamma (k+n)}{\Gamma (2 n+1) \Gamma (-n+k+1)} ( {\gamma}^*)^{2 n} {\dot{\gamma}^*}
\label{nnl6}
\end{eqnarray} 
and the linear terms are $ K_p \gamma^* + \beta_2 {\dot{\gamma}^*}$ which are the (linear) terms of Eq.~(\ref{13b}). The series in Eq.~(\ref{nnl5}) can be summed to give $\epsilon F( \gamma^*,\dot{\gamma}^*)_{nl} = F(  \epsilon_1 \gamma^*,\epsilon_2 \dot{\gamma}^*)_{nl} $ in terms of two separate parameters as
\begin{eqnarray}
\epsilon_1 &=& \beta_1 \left[\sin (\gamma_0) \left(\cos(\gamma_0)+ \beta \sin (\gamma_0) \right)- \left(J_{1}(2\gamma_0)+\beta \pmb{H}_1\left(2  \gamma _0\right) \right) \right],\hspace{2mm}\text{and}\hspace{2mm}\nonumber\\
\epsilon_2 &=& \beta_2 \left[ \left(\cos(\gamma_0)+2 \beta \sin (\gamma_0) \right)- \left(J_{0}(\gamma_0)+2\beta \pmb{H}_0\left(  \gamma _0\right) \right) \right].
\label{nnl7}
\end{eqnarray}  
To the authors' knowledge, the series summations of Weber functions in Eq.~(\ref{nnl5}) are stated here for the first time. Figure~\ref{fig:e1} shows $\epsilon_1/K_p$, part (a), and $\epsilon_2/K_p$, part (b), for TC1. Both are less than unity throughout the range of $\gamma_0$.
The perturbation is independent of $U$ while it increases with $\beta$ and $\sqrt r$. For small $\gamma_0$ the perturbation parameters are independent of $K_p$ and are linear in $\gamma_0$:
\begin{equation}
\epsilon_1 = K_v \left(1- \frac{8}{3 \pi}\right) \gamma_0\hspace{2mm}\text{and}\hspace{2mm} \epsilon_2 = \frac{2 K_v r_u}{\sqrt I_*}  \left(1- \frac{2}{ \pi}\right) \gamma_0.
\label{nnl8}
\end{equation}
The response becomes approximately linear for $\epsilon_2 \ll \beta_2$, or from Eq.~(\ref{nnl8}), when $|\gamma_0| \ll \pi/(2 \beta (\pi-2))$. If a standard value of $\beta=\pi$ is considered representative  of low $\AR$ planforms, the response becomes approximately linear for $|\gamma_0| \ll 25^\circ$. For high $\AR$s with $\beta< \pi$, the linear assumption becomes valid at larger $|\gamma_0|$. Thus the bounding linear solution of Eq.~(\ref{13ccc}) becomes more accurate for high $\AR$s at small $|\gamma_0|$ which is consistent with the yaw response measurements of \citet{barna1976experimental} and \citet{hemsch1990}, who showed that the nonlinear lift decreases with $\AR$. 

It is noted that the ${\beta}_2^2$ term neglected in the solution will become evident only in the second approximation of the solution containing $\epsilon^2$. This justifies neglecting this term in the first approximation of the solution, as stated before.

Another aspect noticed from Fig.~\ref{fig:e1} is that $\epsilon_2>\epsilon_1$ and  $\epsilon_1$ becomes negative for large values of $\gamma_0$ that depend on $\beta$. These indicate an increase in damping and a reduction in damped frequency as $\gamma_0$ increases. This feature is shown in figure 5(a) of \citet{kedr23}, which shows the scaled response measured for two different large values of $\gamma_0$. 
\subsection{Truncation error estimation and improved estimate for the frequency }
The error in truncating the Chebyshev series at $m=2$ will increase as the amplitude $a$ reduces as $t$ increases. The error comes mainly from series representation of the $|\sin \gamma|$ terms in (\ref{5}). The truncation error of the Chebyshev approximation to $|\sin y|$ has a maximum value at $y=0$ and then decreases exponentially with $y$, see \citet{Rouba2018}. The error  is also a function of $\gamma_0$ as given by the coefficients of the series in terms of ${_2}\tilde{ F}_3$ in (\ref{eqnH}). The error will be higher in  $F_1(a)$ in Eq.~(\ref{F1}) than in $G_1(a)$ in (\ref{G1}) and consequently will affect the determination of $\nu$  at large time. The expression for ${\nu}^2$ in Eq.~(\ref{eqn5}) has expansion coefficients $ a_n= (-1)^n (2n+1) \,{_2}\tilde{ F}_3\left(2,1;5/2,n+5/2,3/2-n;- \gamma _0^2\right)$ that converge rapidly as $n$ gets larger so the error can be approximated as $e_n \leq |a_{m+1}|$, \citet{cody1970}. For $m=2$ the error from Eq.~(\ref{eqn5}) is
\begin{equation}
e_n \leq -\frac{14 {\beta}_1  {\beta}|\gamma _0|\gamma _0}{ \sqrt{\pi} } \,{_2}\tilde{ F}_3\left(2,1;5/2,11/2,-3/2;- \gamma _0^2\right)
     \label{e9}
\end{equation}
This error oscillates $m+1$ times over the interval $a \in [0,1]$.  For the TCs, $e_n$  will change sign twice but  decreases exponentially.  $e_n$ was subtracted from ${\nu}^2$ in Eq.~(\ref{eqn5}) to improve the estimation at large $t$ without significantly compromising the calculations at small $t$.  
\section{Analytical solution of yaw response}
\label{sec:analytical}
The coefficients ${A_k^{(1)}}$ and  ${B_k^{(1)}}$ of the differential equations (\ref{27}, \ref{28}) were obtained from  Eqs.~(\ref{25} and \ref{26}). Analytical solutions to   Eqs.~(\ref{27}, \ref{28}) are found in \citet{Srivastava1989}. The solution for $a$ in  Eq.~(\ref{27}) is
\begin{equation}
 a= \frac{a_0 \exp(-\sigma t)}{ \sqrt {1- {A_3}^{(1)}{a_0}^2\left(1-\exp(-2\sigma t)\right)/\sigma- A_5^{(1)}{a_0}^4\left(1-\exp(-4\sigma t)\right)/2\sigma}}
   \label{34}
\end{equation}
 Similarly, the solution to $\psi$ in Eq.~(\ref{28}) is
\begin{equation}
 \psi= \omega_0 t + B_3^{(1)} {K}_1 a_0^2+ B_5^{(1)} {K}_2 a_0^4 +\psi_0.
   \label{35}
\end{equation}
The expressions for ${K}_1$ and ${K}_2$ are 
\begin{align}
 {K}_1=& -\frac{1}{2\sigma \sqrt{-S}}\\ &\times\log\Bigg| \frac{S+Q^2+2R\left[Q\left(1+\exp(2\sigma t)\right)-\sqrt{-S}\left(1-\exp(2\sigma t)\right)\right]+4R^2\exp(2\sigma t)}{S+Q^2+2R\left[Q\left(1+\exp(2\sigma t)\right)+\sqrt{-S}\left(1-\exp(2 \sigma t)\right)\right]+4R^2\exp(2\sigma t)}\Bigg|
   \label{36}
\end{align}
 and
\begin{equation}
 {K}_2= \frac{1}{2\sigma S} \left[\frac{2P+Q\exp(2\sigma t)}{P+Q\exp(2\sigma t)+R\exp(4\sigma t)}-\frac{2P+Q}{P+Q+R}\right]-\frac{Q}{S}{K}_1
   \label{37}
\end{equation}
 The coefficients $S$, $Q$, and $R$ and $P$, as given by \citet{SrivastavaErr1991}, are
\begin{equation}
 S=\frac{a_0^4}{\sigma}\left[2\sigma A_5^{(1)}-A_3^{(1)}-2A_3^{(1)}A_5^{(1)}a_0^2-{A_3^{(1)}}^2a_0^4\right],
   \label{38}
\end{equation}
\begin{equation}
P=1-\frac{A_3^{(1)}}{\sigma}{a_0}^2-\frac{{A_5}^{(1)}}{2\sigma}a_0^4,
   \label{39}
\end{equation}
\begin{equation}
Q=\frac{A_3^{(1)}}{\sigma}a_0^2,
   \label{40}
\end{equation}
and
\begin{equation}
R=\frac{A_5^{(1)}}{2\sigma}a_0^4.
   \label{41}
\end{equation}
The constants $a_0$ and $\psi_0$ are determined from the initial conditions as follows:
\begin{equation}
-\dot{\psi}(a_0) a_0 \sin {\psi_0}+\dot{a}(a_0) \cos {\psi_0} = 0
\label{42}
\end{equation}
and
\begin{equation}
\psi_0= \arccos(1/a_0).
\label{43}
\end{equation}
For $a_0 \approx 1$ as  is usually the case:
\begin{equation}
a_0=1+\frac{1}{2} \left[\frac{A_3^{(1)}+A_5^{(1)}-\sigma}{B_3^{(1)}+B_5^{(1)}+{\omega}_0} \right]^2.
\label{eqa0}
\end{equation} Hence, ${\psi}_0$ is determined from Eq.~(\ref{43}).
\section{Numerical validation of the yaw response solution}
\label{sec:calculations}

\begin{figure}
  \centering
     \begin{subfigure}[b]
     {0.49\textwidth}
         \centering
       \includegraphics[width=\textwidth]{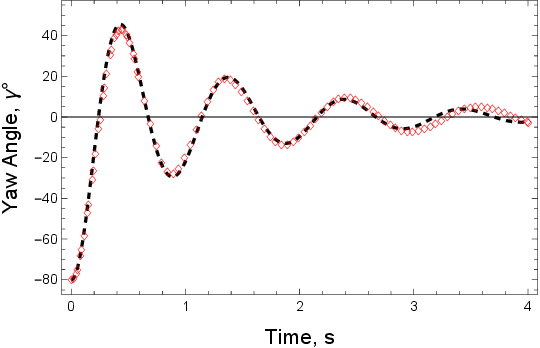}
       \caption{$\beta=3.45$, $\epsilon_1 =-0.144$, and $\epsilon_2 =0.245$.}
     \end{subfigure}
      \begin{subfigure}[b]
      {0.49\textwidth}
         \centering
       \includegraphics[width=\textwidth]{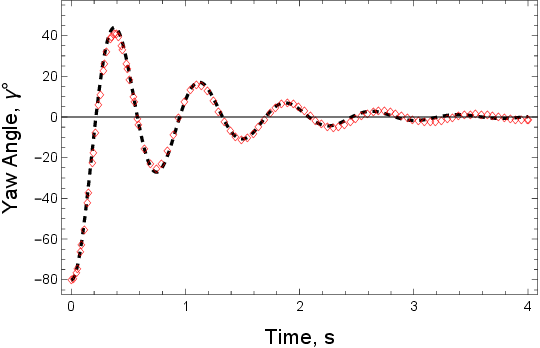}
       \caption{$\beta=1.43$, $\epsilon_1=-0.365$, and $\epsilon_2=0.127$.}
     \end{subfigure}
     \caption{Results for a) TC1, b) TC2 at $U = 17$ m/s and  $\gamma_0=-80^\circ$ with different $\beta = K_v/K_p$. Numerical solution in red diamonds and analytical solution in black dashed lines.} 
\label{fig:TC}
\end{figure}

We now compare the analytical solution of  $\gamma={\gamma}_0 a \cos{\psi}$, where $a$ and $\psi$ are given by Eqs.~(\ref{34} and \ref{35}) respectively, to the numerical solution of Eq.~(\ref{4})  without the ${r}_u^2$ term as discussed above.  Figure~\ref{fig:TC}(a) compares the analytical and numerical solutions for TC1 defined in Subsection \ref{simp}, with $\gamma_0=-80^\circ$ and $U=17 $ m/s  for which $\epsilon_1 =-0.144$ and $\epsilon_2 =0.245$  from Eq.~(\ref{nnl7}) when $\beta = 3.45$. (Figure~\ref{fig:TC}(b) is discussed in the last paragraph of this section). The analytic solution is accurate except at large $t$ where the frequency deviates from the numerical calculations.  The truncation error either changes sign or becomes much smaller than the value predicted by Eq.~(\ref{e9}) for this range. Even when $\epsilon  > 1$, however, nonlinear functions that can be expressed as polynomials yield perturbation solutions that converge to their fundamental periodic solution for $\gamma^*_0$ in Eq.~(\ref{15}). Thus, the method will prevent the error in frequency and damping from transferring to the higher order $\epsilon$ terms, \citet{senator1993}. The KBM method was shown to provide accurate solutions for  $\epsilon=1$ by, for example, \citet{Mendelson1970},  using only the fundamental harmonic of the $\epsilon^0$ solution. This means that the deviation in the analytical solution in Fig.~\ref{fig:TC}(a) results mainly from truncating the nonlinear function to allow analytical integration of $a$ and $\psi$. At the same time, the first order perturbation analysis is sufficient to obtain an accurate solution for  TC1 as discussed above and as will be demonstrated further in the next section.
Another aspect that was not anticipated when this study began is   the  \textit{decrease} in the nonlinearity and the  series truncation error with increasing $\beta= K_v/K_p$ as discussed in the previous section.  The vortex lift results partly from the loss of potential flow due to the suction force resulting from flow separation at the leading edge, which is the basis of the LESA of \citet{Polhamus1966}.  He gives $K_v$ as 
\begin{equation}
   K_v= K_p \left(1-K_p K_i\right)/\cos \Lambda 
  \label{kv}
\end{equation}
where $K_i =\partial {C_{Di}}/\partial {{C}_l^2}$.  The induced drag coefficient $C_{Di}$ due to the lift coefficient ${C}_l$  can be approximated as $K_i \approx1/(\pi \AR)$ from potential flow theory. In contrast,  the LESA implies the complete loss of leading edge thrust and a fully developed vortex vortex flow which gives  ${C_{Di}}/{{C}_l^2}= \tan \alpha / C_l$, \citet{Polhamus1971}. This means that $K_v$ changes significantly with $\AR$ as well as with $\alpha$. \citet{Polhamus1966} found that $K_v$ changes very little from the value of $\pi$ with $\AR$ which according to \citet{traub2018} is responsible for inaccurate prediction of $C_l$ using LESA at high $\AR$.  \citet{traub2018} derived a simple correlation for $K_v$ that shows a decrease in $K_v$ with $\AR$ to asymptote to $\pi$ as $\AR \downarrow 0$. The $K_v$ dependence on $\alpha$ was derived by \citet{purvis1981}.  $K_v =2.36$ was found for delta tail fin of $\AR=1.97$ in \citet{kedr25} using system identification for TC response, a value that is very close to Traub's \citeyearpar{traub2018} correlation. Hence, $\epsilon$ will have lower values than given in Fig.~\ref{fig:e1} as the $\AR$ increases as discussed above and in Section \ref{sec:epsilon}. In addition, the truncation error will be lower as can be shown by Eq.~(\ref{e9}). 

Because the experimental implementation of the TC is very straightforward, we believe that the force coefficients, $K_p$ and $K_v$, are best obtained using system identification as discussed in \citet{kedr25}, whereas any change with the $\gamma$ is dealt with using the separation functions mentioned in Section \ref{sec:headings} and discussed further in Section \ref{sec:coeff}.  This is especially the case for the complicated geometries that some small wind turbine manufacturers use for their tail fins, an example of which is described and tested in  \citet{kedr25}. 

Figure \ref{fig:TC}(b) shows the response of test case TC2  with a delta planform but with $c_0= 0.143$ m and $b_0=0.141 $ m, as given in \citet{kedr23}. It has a relatively large  $\AR = 1.97$, for which  the theoretical values of $K_p= 2.184 $ and $K_v=\pi$ are used, $r=0.5383$ m, $I= 0.0376 $~kg m$^2$, \citet{kedr23}, $\gamma_0=-80^\circ$, and $U=17 $ m/s, giving  $I_* = 0.04$ s$^2$. $\beta = 1.43$ which is less than half that of TC1. Thus, Eq.~(\ref{nnl7}) gives $\epsilon_1 =-0.365$ and $\epsilon_2 =0.127$. The comparison with the numerical solution shows, as expected, that the prediction at large $t$ is better for this higher $\AR$ planform. 

\section{Accurate solution of the minimal equation via the Beecham and Titchener method}
\label{sec:ARC}
A more precise solution of the minimal equation can be obtained using the BT method, \citet{BEECHAM1971}, which  avoids the constraints applied in Eq.~(\ref{n1}) and  the error in expressing the nonlinear functions as truncated polynomials. These advantages of the BT over the KBM method also occur for the equations analysed by \citet{Christopher1985}. It does not, however,  give the amplitude ($a$) and phase ($\psi$) in  closed form   except for the limiting cases. 

The restriction used to derive the rate of change of the response in Section \ref{sec:solution} is discussed in detail in \citet{Simpson1977}. The restriction is necessary to formulate $\gamma$ in terms of  $a$ and $\psi$. This is in accordance with the method of variation of parameters and constraints of the form 
\begin{equation}
 g (a, \zeta, \omega) \cos \psi+ h(a,\zeta, \omega) \sin \psi=0
   \label{6_1}
\end{equation}
where $g,h$ are functions of the two variables $\zeta$ and $\omega$ expressed by Eqs.~(\ref{19} and \ref{20}), respectively. In the KBM  the constraints allow formulating $\dot{\gamma}^*$ in a form consistent with the linear response of a second order oscillating system. In this section, however, we use the BT method with its fewer  constraints on the definition of $\dot{\gamma}^*$.  This results in more accurate, but  more complicated, expressions for the $a$ and $\psi$  which do not have closed forms. The BT method, however, provides a compact form of the solution that is better for the analysis of the yaw response.

The BT method uses the two following parameters: $\lambda = \dot{a}_r/ a_r$ and the frequency ${\omega}_r^2$,  so that ${\omega}_r= \dot{\psi}_r$, where the subscript $r$ is used distinguish them from those obtained using the KBM method. The two parameters are given in two stage approximations. In the first approximation their definitions are the same as the KBM method but without restriction on the definition of $\dot{\gamma}^*$, and the variables are
\begin{equation}
 \lambda_1= I_1 /(2\omega_{r,1} )
   \label{6_2}
\end{equation}
where the subscript $1$  indicates the first approximation. In this approximation, the time rate of change of $\lambda$ and $\omega_r$ in the definition of ${\ddot{\gamma}^*}_r$ are disregarded. Similarly, $\omega_{r,1}$ is given by
\begin{equation}
 {\omega}_{r,1}^2= {\lambda}_1^2 + I_2 
   \label{6_2x}
\end{equation}
where $I_1$ and $I_2$  are the averages of the nonlinear $F$ function, defined in Eq.~(\ref{7}), over one period of oscillation. They are equivalent to $G_1$ and $F_1$ respectively of the KBM method in Eqs.~(\ref{33} and \ref{32}): 
\begin{eqnarray}  
  I_1&=& \frac{1}{\pi a_r}\int_0^{2\pi} {F({\gamma}^{*}_r,\dot{\gamma}^{*}_r})   \sin {\psi_r} d{\psi_r} \nonumber\\ I_2&=& \frac{1}{\pi a_r}\int_0^{2\pi} {F({\gamma}^{*}_r,\dot{\gamma}^{*}_r})  \cos {\psi_r} d{\psi_r}
   \label{6_3}
\end{eqnarray}
where ${\gamma}^{*}_r$ is defined similarly to KBM method as
\begin{equation}
 {\gamma}^{*}_r= a_r \cos{\psi}_r.
   \label{6_4}
\end{equation}
The assumption used to define $\dot{\gamma}^{*}$ in Eq.~(\ref{n2}) is revised to give
\begin{equation}
 \dot{\gamma}^{*}_r= \lambda_1  {a_r} \cos{\psi}_r - \omega_{r,1} {a_r} \sin{\psi}_r
   \label{6_5}
   \end{equation}
Then, the first approximations for $\lambda_r$ and $\omega_r$ are used to find more accurate expressions for  ${\ddot{\gamma}^*}_r$ and avoid the assumptions of the first approximation. For the second approximation, \citet{BEECHAM1971} give
\begin{equation}
\lambda_2= \frac{I_1}{2 \omega_{r,1} \left[1+\frac{a_{r}}{4  {\omega}_{r,1}^2} \frac{d {\omega}_{r,1}^2 }{d a_{r}}\right]}
   \label{6_6}
\end{equation} and
\begin{equation}
{\omega}_{r,2}^2= {\lambda}_1^2+ I_2+ {\lambda}_1 a_{r} \frac{d {\lambda}_1}{d a_{r}}. 
   \label{6_7}
\end{equation} 
This use of the first approximation to the parameters in the definitions of $\lambda_2$ and ${\omega}_{r,2}$ represents the main assumption of the BT method, \citet{Simpson1977}. The consequences of this assumption depends on the particular form of the nonlinear function $F$; for the current study we show below that the resulting error is negligible.

\begin{figure}
\centerline{\includegraphics{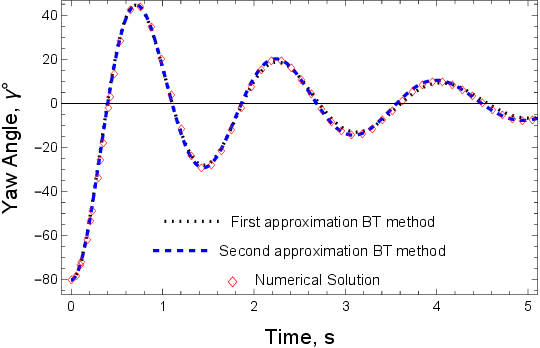}}
  \caption{Results for the BT method for TC1 at $U = 10 $m/s. Numerical solution of Eq.~(\ref{4}) in  red diamonds, BT first approximation  in dotted black, while second approximation for a delta tail fin  released at $\gamma_0=-80^\circ$.}  
\label{fig:RAE}
\end{figure}

The evaluation of $I_1$ and $I_2$ in Eq.~(\ref{6_3}) were obtained using Mathematica. The advantage of the BT method is that no perturbation analysis has to be done after averaging. Thus, the BT method can incorporate the ${\beta}_2^2$ term that is neglected in the KBM solution.  The first approximations,  again neglecting terms of order $\beta_2^3$ and higher as discussed in subsection \ref{simpx}, are
 \begin{eqnarray}  
  I_1= -\frac{2 \beta _2 \omega_{r,1}  J_1\left(\gamma_0  a_r\right)}{a_r \gamma_0 }-\frac{4 \beta  \beta _2 \cos \mu \sign\left(\gamma _0\right) \pmb{H}_1\left(\gamma_0  a_r\right)}{{a}_r^2 \gamma_0 }-\frac{8 \beta {\beta} _2^2 \lambda_1  \omega_{r,1}  a_r}{3 \pi  \beta _1}, 
     \label{6_8}
\end{eqnarray}
and
\begin{eqnarray} 
I_2&=& \beta _2 \lambda_1  \left(J_0\left(\gamma_0  a_r\right)-J_2\left(\gamma_0  a_r\right)\right)+\frac{\beta _1 J_1\left(2 \gamma_0  a_r\right)}{a_r}\nonumber\\&& +\frac{4 \beta  \beta _2 \sin \mu  \sign\left(\gamma _0\right) \left( a_r \gamma_0  \pmb{H}_0\left(\gamma_0  a_r\right)-  \pmb{H}_1\left(\gamma_0  a_r\right)\right)}{{a}_r^2 \gamma_0 }+\frac{\beta \beta _1  \sign\left(\gamma _0\right)\pmb{H}_{1}\left(2 \gamma_0  a_r\right)}{  a_r}\nonumber\\&&+\frac{4 \beta  {\beta}_2^2 a_r \left(2 {\lambda}_1^2+{\omega}_{r,1} ^2\right)}{3 \pi  \beta _1}
 \label{6_9}
\end{eqnarray}
where $\sin\mu= \lambda_1 a_r$, $\cos\mu= a_r \omega_{r,1}$. This gives $\lambda_1$ as
\begin{eqnarray}  
 \lambda_1 =-\frac{3 \pi  \beta _1 \beta _2 \left(J_1\left(\gamma_0  a_r\right)+2 \beta \sign\left(\gamma _0\right) \pmb{H}_1\left(\gamma_0  a_r\right)\right)}{a_r \gamma_0  \left(4 \beta  {\beta}_2^2 a_r+3 \pi  \beta _1\right)}.  
     \label{6_10}
\end{eqnarray}
$\omega_{r,1}$ is obtained from Eq.~(\ref{6_2x}) as
\begin{eqnarray}  
{\omega}_{r,1}^2 &=& \frac{3 \pi  {\beta}_1^2 \left(J_1\left(2 a_r \gamma _0\right)+\beta \sign\left(\gamma _0\right) \pmb{H}_1\left(2 a_r \gamma _0\right)\right)}{a_r \left(4 \beta  {\beta}_2^2 a_r+3 \pi  \beta _1\right)}\nonumber\\&&+\frac{\beta _1 {\beta}_2^2}{{\gamma}_0^2 {a}_r^2 \left(4 \beta  {\beta} _2^2 a_r+3 \pi  \beta _1\right)}\Bigg[3 \pi  \left(J_1\left(a_r \gamma _0\right){}^2-4 \beta ^2 \pmb{H}_1\left(a_r \gamma _0\right){}^2\right)\nonumber\\&&+ 8\beta  {\gamma} _0^2 {a}_r^2 \big(-J_1\left(a_r \gamma _0\right)+J_1\left(2 a_r \gamma _0\right)-2 \beta  \sign\left(\gamma _0\right)\pmb{H}_1\left(a_r \gamma _0\right) \nonumber\\&&+\beta \sign\left(\gamma _0\right) \pmb{H}_1\left(2 a_r \gamma _0\right)\big)+3 \pi  \gamma _0 a_r \left(J_1\left(a_r \gamma _0\right)+2 \beta \sign\left(\gamma _0\right) \pmb{H}_1\left(a_r \gamma _0\right)\right)\nonumber\\&& \times\left(-J_0\left(a_r \gamma _0\right)+J_2\left(a_r \gamma _0\right)+4 \beta \sign\left(\gamma _0\right) \pmb{H}_2\left(a_r \gamma _0\right)\right)\bigg].  
     \label{wr}
\end{eqnarray}
The expressions for $\lambda_2$ and $\omega_{r_2}$ are too complicated to state here. The Mathematica scripts for their evaluation are available on request from the authors.  Nevertheless, both the first and second approximations  will be compared against the numerical solution.  The response  $\gamma^* =  a_r \cos {\psi_r}$ using the first and second approximations are compared in Fig.~\ref{fig:RAE} to the numerical solution of the TC1  for $U= 10$ m/s. 

The agreement between the first and second BT solutions and the numerical solution is very good. The first approximation shows similar accuracy at high angle and deviates slightly from both the more accurate second approximation and the numerical solution at large $t$ and small $\gamma$. This demonstrates that accuracy of the averaging method; the small deviation of the analytical solution in Fig.~\ref{fig:TC} comes from the truncation error so the ${\gamma}^{*}_0$ solution in Eq.~(\ref{16}) is sufficient. Also, it shows that approximating $\dot {\gamma}^{*}$ by Eq.~(\ref{n2}) is justified and this demonstrates the validity of the KBM method for the minimal equation.

The accuracy of the first-order approximation and the more compact form of the $\lambda_1$ and ${\omega}_{r,1}$ make them convenient for analyzing the different aspects of the solutions in the following sections. To make the analysis clearer, the neglect of the $\textit{O}({\beta}_2^2)$ terms as done in the KBM method, gives the following simple expressions for $\lambda_1$ and ${\omega}_{r,1}$ with the subscript ``1"  dropped:
\begin{eqnarray}  
 \lambda =- \beta _2  \left[J_1\left(\gamma_0  a_r\right)+2 \beta \sign\left(\gamma _0\right) \pmb{H}_1\left(\gamma_0  a_r\right)\right]/(a_r \gamma_0),  
     \label{lam1}
\end{eqnarray}
and
\begin{eqnarray}  
{\omega}_r^2 &=& \beta _1 \left[J_1\left(2 a_r \gamma _0\right)+\beta \sign\left(\gamma _0\right) \pmb{H}_1\left(2 a_r \gamma _0\right)\right]/a_r.  
     \label{wr1}
\end{eqnarray}
The expressions are comparable to  $G_1$ and $F_1$ in Eqs.~(\ref{33s} and \ref{32s}) respectively. It also shows that the assumption is equivalent to neglecting ${\lambda}_1^2$ from the expression of ${\omega_{r,1}}$ in Eq.~(\ref{6_2x}).
\subsection{Equivalent linear response}
\label{equivalent linear}
An advantage of the averaging method is the possibility of obtaining a linear system equivalent to the nonlinear one. and so defining the  equivalent damping ratio and frequency. Those parameters are valuable in control design and system identification, e.g. \citet{barton1965}. The physical interpretation of the ``equivalent'' system is that both systems have the same energy in a period of oscillation, see  \citet[Ch.14, Sec.7]{Minorsky62}.

\begin{figure}
\centerline{\includegraphics{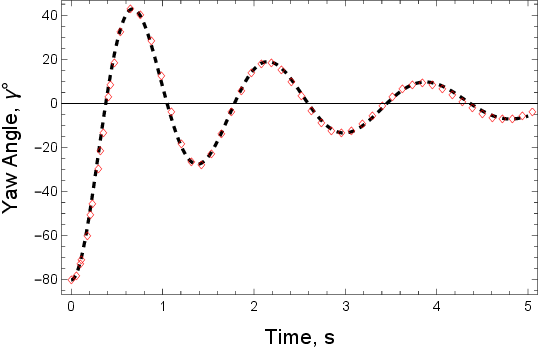}}
  \caption{Equivalent linear solution for TC1 at $U = 10 $m/s. Numerical solution of Eq.~(\ref{5}) in red diamond, equivalent linear solution of Eq.~(\ref{6_11}) in dashed black of yaw response of a delta fin  released at $\gamma_0=-80^\circ$.}  
\label{fig:eq}
\end{figure}

An equivalent linear response in the form 
\begin{equation}
\ddot{\gamma}^*_r + 2{\sigma_e} \dot{\gamma}^{*}_r+ {\omega_e^{2}} {\gamma}^{*}_r=0 
   \label{6_11}
\end{equation} 
can be obtained using the above analysis, so that ${\sigma_e}=-\lambda$ and ${\omega_e}= \sqrt{{\omega_r^{2}}+ \lambda^2} \approx \omega_r$, \citet{BEECHAM1971}.
To check the validity of the equivalent system, Eq.~(\ref{6_11}) is solved numerically and compared to to the numerical solution of Eq.~(\ref{5}) with the ${\beta}_2^2$ term  ignored. Figure \ref{fig:eq} shows a satisfactory comparison which demonstrates the accuracy and hence the efficacy of the equivalent linear system. It also validates the  approximation  $\omega_e \approx \omega_r$.
The equivalent system damping ratio, $\zeta_e$, is given by   
\begin{equation}
\zeta_e= \frac{\sigma_e}{\omega_e}=\frac{\beta _2}{\sqrt{\beta _1} \gamma_0 } \frac{ J_1\left(\gamma_0  a_r\right)+2 \beta \sign\left(\gamma _0\right) \pmb{H}_1\left(\gamma_0  a_r\right)}{ \sqrt{a_r \left(J_1\left(2 \gamma_0  a_r\right)+\beta  \sign\left(\gamma _0\right)\pmb{H}_1\left(2 \gamma_0  a_r\right)\right)}}
   \label{6_12}
\end{equation} 

\subsection{Logarithmic decrement and response envelope}
\label{LD}
\begin{figure}
\centerline{\includegraphics{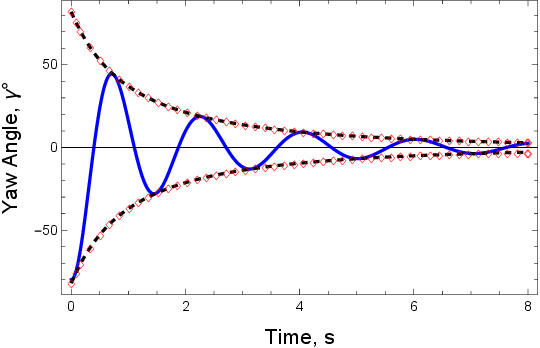}}
  \caption{Response envelope by considering the $\sqrt{1+{\zeta}_e^2}$ factor in red diamonds and by disregarding it shown in dashed black for TC1 at $U = 10$ m/s, while the full response is shown in solid blue.}  
\label{fig:env}
\end{figure}

The logarithmic decrement is a significant parameter in characterising the damping. It is usually linked to linear systems, but when the damping is small, as in the current case, the concept can be used without serious error, \citet{Rasmussen1977}.
The yaw rate is given exactly by Eq.~(\ref{6_5}). Setting $\dot{\gamma}^*=0$ gives the equation for the extrema, ${\gamma}^{*}_p$, as
\begin{equation}
 {\gamma}^{*}_p= a_p /\sqrt{1+\left(\lambda/\omega_r\right)^2}=a_p/ \sqrt{1+{\zeta}_e^2} \approx a_p
   \label{l1}
\end{equation}
where $a_p$ is the amplitude at ${\gamma}^{*}_p$. The relation shows that there is a small  phase shift between  ${\gamma}^{*}_p$ and the points of coincidence of $\gamma^*$ and the response envelope. By neglecting the ${\beta}_2^2$ terms of the solution,  however,  they coincide. The exact relation for the logarithmic decrement, $\delta$, of any cycle $i$  is
\begin{equation}
\delta  =    \log (a_i/a_{i+1}) \approx \Delta a/a,
   \label{l2x}
\end{equation}
 \citet{Panovko1971}. The approximation is valid for moderate damping where the amplitude changes slowly in one cycle.  $\Delta a =T da/dt$ where $T=(2 \pi)/ \omega_e$ is the period of oscillation. Thus, $\delta$ is given by
\begin{equation}
\delta= - 2 \pi \lambda/\omega_e= 2 \pi \zeta_e.
   \label{l2}
\end{equation}
For moderate damping, therefore, the logarithmic decrement is equivalent to the damping ratio as in linear systems but is a function of the amplitude. The response envelope from Eq.~(\ref{l1}) is plotted in Fig.~\ref{fig:env} with $a$  obtained to similar accuracy from either the KBM method, Eq.~(\ref{34}), or by numerically integrating $\lambda$ from Eq.~(\ref{lam1}). The figure shows the envelope by considering the factor $\sqrt{1+ {\zeta}_e^2}$ and by neglecting it.  The effect of $\zeta_e$ on the response envelope is negligible and Eq.~(\ref{l2}) holds with high accuracy. 

\section{ Beecham and Titchener analysis and  corresponding limiting models}
\label{sec:analysis}
The BT method has the advantage of providing a compact, accurate expression for the response in terms of the two parameters characterising the nonlinear response, namely  $\lambda$ and $\omega_r$. These parameters  include the  ${\beta}_2^2$ terms, whereas the KPM method uses the $\textit{O}(\epsilon^2)$ approximation, given by (\ref{6_10} and \ref{wr}) for $\lambda$ and $\omega_r$ respectively. In addition, it was shown in Section \ref{sec:epsilon} that $\epsilon$ is predominantly linear in $\gamma_0$ for an appreciable range of  $\gamma_0$. In fact, the actual response is controlled by  $K_p$, $K_v$, and $C_{Dc}$ at  large $\gamma_0$, \citet{kedr23}. The determination of  these constants is vital for the successful system identification of any tail fin response,  \citet{kedr25}. In this section, therefore, two important limiting cases solutions are developed.   We will show that, unexpectedly, $K_v$ enters the small$-|\gamma|$ response equation.

\subsection{Analytical solution for small yaw angle}
\label{sec:small}

\begin{figure}
\centerline{\includegraphics{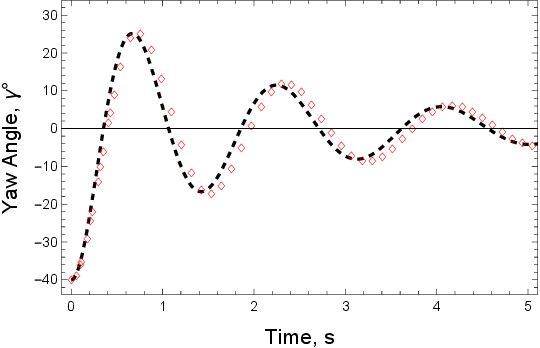}}
  \caption{Results for TC1 at $U = 10$ m/s. Analytical solution of the low $\gamma_0$ in dashed black compared to the numerical solution in red dimaonds  at $\gamma_0=-40^\circ$.}  
\label{fig:lin1}
\end{figure}

An analytical solution for very small $|\gamma_0|$ was derived in Section \ref{bound}. This second order, linear solution neglects the $K_v$ term in Eq.~(\ref{3}), and was shown to apply for $|\gamma_0| \ll 25^\circ$ for a typical  fin in Section \ref{sec:epsilon}. This is the basic model used in characterising wind vane response, see for example \citet{barna1976experimental} and \citet{Kris94}.  \citet{barna1976experimental}, however, found the model failed to  predict accurately the damping and frequency of their wind vanes, especially those with low $\AR$, even for $|\gamma_0|$ as low as $5 ^\circ $. In Section \ref{sec:epsilon},  this failure was attributed to  high values of $\beta = K_v/K_p$ and the consequent increase of nonlinearity.  By also using Eq.~(\ref{13b}), \citet{kedr25} estimated values of $K_p$ higher than calculated from potential flow theories.  In contrast, the solution we now derive by keeping the leading terms of both  potential and vortex flow can be used to identify \textit{both} $K_p$ and $K_v$. In addition, the new solution will extend the upper bound of the linear solution.
Only the first order Eqs.~(\ref{lam1},~\ref{wr1}) are considered.  For small $|\gamma_0|$, the leading terms of $J_1(\gamma_0  a_r) \sim \left(\gamma_0  a_r \right)/2$, and $\pmb{H}_1\left(\gamma_0  a_r\right) \sim 2{\left(\gamma_0  a_r \right)}^2/(3 \pi) $. The equation for $a_r$ becomes
\begin{eqnarray}  
\frac{1}{a_r} \frac{d a_r}{ dt} =- \frac{\beta _2}{2}    \left(1+\frac{8}{3 \pi} \beta |\gamma _0| a_r \right)  
     \label{smlam1}
\end{eqnarray}
which is easily separated and factored to give
\begin{equation}
 a_r =  a_{r0}/\big[ \exp(\beta _2 t/2)+8\beta | \gamma _0| a_{r0} \big(\exp(\beta _2 t/2)-1\big)/(3\pi)\big].
     \label{smlam2}
\end{equation} where $a_{r0}$ is the value of $a_r$ at $t=0$. 
Equation (\ref{wr1}) becomes
\begin{equation}  
\frac{d\psi_r}{dt}={\omega}_r= \sqrt{\beta_1 \gamma_0  \left(1+8 \beta |\gamma _0| a_r /(3\pi)\right)} .  
     \label{wrsm1}
\end{equation}With the help of Eq.~(\ref{smlam1}), $\psi_r$ is given by
\begin{equation}  
\psi_r-\psi_{r0}=-\frac{2 \sqrt{\beta_1 \gamma_0} }{\beta_2} \int_ {a_{r0}}^ {a_r} \frac{d a_r}{a_r \sqrt{ \left(1+8 \beta |\gamma _0| a_r /(3\pi)\right)}}.  
     \label{wrsm2}
\end{equation}
The solution for the integral, \citet[No. 2.224.5]{gradshteyn2014table}, is
\begin{equation}  
\psi _r=\psi _{{r0}}+\frac{1}{ \zeta_{\text{lin}}}\log \left| \frac{\left(\sqrt{8 \beta |\gamma _0| a_r /(3\pi)}+1\right) \left(\sqrt{8 \beta |\gamma _0| a_{r0}/(3\pi)}-1\right)}{\left(\sqrt{8 \beta |\gamma _0| a_r /(3\pi)}-1\right) \left(\sqrt{8 \beta |\gamma _0| a_{r0} /(3\pi)}+1\right)}\right|   
     \label{wrsm3}
\end{equation}
where $ \zeta_{\text{lin}}= \beta_2 /(2 \sqrt{\beta_1 \gamma_0})$, the  damping ratio from Eq.~(\ref{6_12}), is the same as the linear result in Eq.~(\ref{13cc}), and $\psi _{{r0}}$ is $\psi_r$ at $t=0$.  The $K_v-$dependency comes from the logarithmic term. The initial conditions ${\gamma}^{*}_r(0)=1, \dot{\gamma}^{*}_r(0)=0$ give $a_{r0}$ and $\psi _{{r0}}$ from Eqs.~(\ref{6_4} and \ref{6_5}) as
\begin{equation}  
a_{r0}= \left(K^*+\sqrt{{K^*}^2+4(1+{\zeta}_{\text{lin}}^2)}\right)/2\approx \left(K^*+2\right) /2 
     \label{smlam3}
\end{equation}
where $K^*=8 \beta |\gamma_0|\zeta_{\text{lin}}^2/(3 \pi)$ is  also a function of $K_v$ and $a_{r0} =1$ only in the limit $K_v\downarrow 0$.  Further,
\begin{equation}  
\psi _{{r0}}= \arccos(1/a_{r0}).  
     \label{wrsm4}
\end{equation}
A bound for the validity of this small $|\gamma_0|$ approximation of  Eqs.~(\ref{smlam1} and \ref{wrsm3}) is set by the $J_1(2 \gamma_0)$ approximation in the expression for $\omega_r$ in Eq.~(\ref{wr1}) to give the limit of $J_n(2 \gamma_0) \rightarrow \gamma_0$ as $ 0< 2\gamma_0 \ll \sqrt{1+n}$, or $|\gamma_0| \ll 180/(\sqrt{2}{\pi}) \approx 40.5^ \circ$. 
To verify this, this analytical solution is compared with the numerical solution of Eq.~(\ref{5}) for $\gamma_0 = -40 ^ \circ$ in Fig.~\ref{fig:lin1}. The comparison is sufficiently satisfactory for the low$-|\gamma_0|$ solution to be accurate  for $|\gamma_0| < 40.5^\circ$.  The deviation in predicted frequency arises from using the small angle approximation for $\omega_r$ in Eq.~(\ref{wr1}) because it has $2 \gamma_0$ as the function argument. 

\citet{kedr23}  found that the measured response in terms of $\gamma^*$ changes with $|\gamma_0|$. We again use the subscript ``$p$'' to denote an extrema, in this case, the subsequent one. Their figure 5 shows results for $\gamma_0 =-80^\circ$. For $|\gamma| \le \gamma_p = 46.82^\circ$, that is with $\gamma_0$ replaced by $\gamma_p$ and $t$ replaced by $t-t_p$,  the data plotted as $\gamma^*$ were coincident with those for $\gamma_0 =-40^\circ$ where the low$-|\gamma_0|$ approximation should hold.  This result provides a formula for the time shift, $t_p$, from $\gamma_{0}$ to the next $\gamma_{p}$. The only requirement for collapse is that the amplitude $\gamma_0 a_r$ is the same for the two response or that $a_r \propto \gamma_{p}/\gamma_0$.
From Eq.~(\ref{smlam2}) we have
\begin{equation}  
t_{p}= \frac{2}{\beta_2}\log \left| \frac{ a_{{r0}} \left| \gamma_0\right| \left(1+8 \beta  \left| \gamma_p\right| a_{{rp}}/(3 \pi) \right) }{ a_{{rp}}  \left| \gamma_p\right|\left(1+8 \beta  \left| \gamma _0\right|  a_{{r0}}/(3 \pi)\right)}\right|  
     \label{tp1}
\end{equation}
where $a_{{rp}}$ is the amplitude at $\gamma_p$.  In the case of Fig.~\ref{fig:lin1}, for example, the time to reach the peak $\gamma_{p1}= 25.1 ^\circ$ calculated using Eq.~(\ref{tp1}) is $0.69 $ s which is exact to two decimal places. 

Using this analysis it can be shown that if it lies within the validity of the current low$-|\gamma_0|$ model, the response is shifted in time from $\gamma_0$ to  $\gamma_p$.  In other words, the response can be considered to restart from $\gamma_p$. The frequency and damping must be equivalent according to the equivalent linear system representation of Section \ref{equivalent linear}. This means that the amplitude is equal to the ratio of the two peaks as stated above and the time shift needed is calculated from Eq.~(\ref{tp1}). Expanding the normalized angle response ${\gamma}^{*}_r= a_r \cos \psi_r$ in the small parameter $g^* = 8 \beta |\gamma_0|/(3 \pi)$, gives to the first order in $g^*$:
\begin{equation}  
{\gamma}^{*}_r=  e^{-K_p r_ut_*/2} \cos \left(\sqrt{K_p}t_*\right)+e^{-K_p r_ut_*} \left(1-e^{-K_p r_ut_*/2}\right) \cos \left( \sqrt{K_p}t_*-\phi _g\right)g^*+ \textit{O}({g^*}^2)
     \label{g1}
\end{equation}
where the first term on the right is the linear response as given by Eq.~(\ref{13ccc}), and $\tan \phi_g= \sqrt{K_p/I_*}/r_u$ is a constant phase shift. If the first bracketed component of the second term 
is substituted from Eq.~(\ref{smlam2}), time is shifted by $\approx2/\beta_2\log{(   \left| \gamma_0\right|/|\gamma_p|)}$, and the amplitude is set to the ratio $a_r=|\gamma_p/\gamma_0|$, then all the $\gamma_0$ in Eq.~(\ref{g1}) will be replaced by $\gamma_p$ giving the same response equation but in terms of the angle $\gamma_p$. The same holds for the terms of higher orders of $g^*$ as well indicating the equivalence of the responses. 
\subsubsection {System identification of the coefficients $K_p$ and $K_v$} \label{ID}
That the amplitude of motion is proportional to the ratio of the angle extrema has an important application to the use of the low$-|\gamma_0|$ equation for system identification  as in \citet{kedr25}. They used Eq.~(\ref{13ccc})  to estimate $K_p$ which ignores the $K_v$ term and thereby overestimated $K_p$ compared to the  values taken from the literature. Subsection \ref{sec:small} showed that at low $\gamma_0$, the extrema of the response can be considered as a new initial condition that is shifted by time $t_p$ from the original one. The test cases TC1 and TC2 are now used to estimate the coefficients $K_p$ and $K_v$ from the low angle section of the original $\gamma_0 \approx -80^ \circ$  data, using Eqs.~(\ref{smlam2}, \ref{wrsm3}, \ref{smlam3}, and  \ref{wrsm4}). Taking $\gamma_p$ now to be any extrema, we arbitrarily chose ones of small magnitude: $\gamma_p = 8.39 ^ \circ$ for TC1 and $-8.75 ^ \circ$ for TC2.  $K_v$ is first optimised and then its definition in terms of $K_p$ given by Eq.~(\ref{kv}) was tested where $K_i$ is estimated instead to check the effect of this on overfitting the coefficients. The function ``FindFit" in Mathematica was used with its default settings. For TC1, $K_p= 0.9$ and $K_v=3.11$ were estimated. These are close to the theoretical values $K_p= \pi \AR/2=0.91 $, and $K_v=\pi$  as quoted above and shown in Table II of \citet{kedr23}. For TC2, the estimated values are $K_p= 2.6$ and $K_v=2.2$; the former is a little higher than the theoretical $K_p= 2.184 $ as quoted above from Table II of \citet{kedr23} and the latter is lower than $K_v=\pi$ from the same source. $K_v$ falling below $\pi$, however, is consistent with the decrease in $K_v$ with increasing $\AR$ as discussed in Section \ref{sec:calculations} and as estimated in \citet{kedr25}. \citet{Wentz1968} note  that according to \citet{Polhamus1966}, the LESA  overpredicts $K_v$ for large $\AR$ delta wings which have insufficient aft surface  to generate full vortex lift. Fitting $K_v$ in terms of $K_p$ was found to give the same values of the estimated coefficients for the two test cases. In addition, the results are independent of the initial setting of the coefficients in ``FindFit" suggesting a global optimum.

For TC1,  $K^*=0.0031$, $\psi_{r0}=-3.3^\circ$, and the $K_v$ term in Eq.~(\ref{wrsm3}) is approximately $ 1.134t$.  For TC2, $K^*=0.0017$, $\psi_{r0}=-2.4^\circ$, and the $K_v$ term $\approx 1.686t$ which shows the smaller dependency on $K_v$ for high $\AR$ tail fins.

\citet{kedr25} also tested and fitted the response of rectangular tail fins. In this case fitting $K_v$ directly using the current model resulted in overpredicted values of $K_v$ compared to those in their study. This is because for rectangular planforms $K_v = K_{v,le}+K_{v,se}$  where ``le'' and ``se'' refer to the leading and side edges respectively. The two components have different dependence on $K_p$ resulting in overfitted values. Alternatively, if  $K_v$ is found in terms of $K_p$ for the two components using Equations (2, 3, and 4) of \citet{traub2023}, $K_p= 0.92$ and $K_v=3.2$ for the low $\AR$  and $K_p= 2.7$ and $K_v=3.6$ for the high $\AR$ one. These values are in very good agreement with the theoretical values in \citet{kedr25} for $K_p$ and the more accurate estimated value of $K_v$. Using the same procedure to fit the two components of $K_v$ separately, gives, for the low $\AR=0.5$ rectangular fin, $K_{v,le}=0.8$ and $K_{v,se}=2.4$, indicating the dominance of the second which asymptotes to $\pi$ as $\AR \rightarrow0$, \citet{devoria2017}. This limit contrasts to the value of 2 obtained in \citet{bollay1939}.  

This discussion and that of Section \ref{sec:calculations} implies that  tail fins and wind vanes need different geometries due to their different modes of operation. Tail fins must be effective at high $|\gamma_0|$ to assist turbine starting at low $U$. \citet{kedr23} note that, once a turbine starts producing power, its tail fin becomes less important for yaw stability. In constrast,  wind vanes require excellent linear behavior to minimise $|\gamma|$ at any $U$.  Hence, it is desirable for wind vanes to have high $\AR$ whose response is dominated by the linear $K_p$ term. For tail fins, the delay of stall is important, which requires a low $\AR$ to give a response dominated by the nonlinear $K_v$  term. It is noted that small wind turbines typically have high $\AR$ tail fins; \citet{kedr25} tested a commercial fin with $\AR = 3.71$.  Apart from varying the planform, the frequency of the response can be altered through  $I_*$ which depends, for example, on the density and thickness of the material comprising the planform and tail boom.
\subsubsection {Independence of initial conditions and requirements for linearity} \label{init}
Equations (\ref{smlam1} and \ref{wrsm1}) suggest  that the low $\gamma_0$ solution can become independent of the initial conditions.  The $\beta-$term for $\lambda$ and $\omega_r$ in Eqs.~(\ref{lam1} and \ref{wr1}),  respectively, contains 
the Struve function $\pmb{H}_n\left(\gamma_0  a_r\right)$ which can be expanded as
\begin{align}  
\pmb{H}_n\left(\gamma_0  a_r\right)&= \frac{-2 {\gamma}^n_0}{\pi} \sin{(\pi \gamma_0)} \sum _{k=1}^\infty    \left[\frac{(-1)^k}{k^2-{\gamma}_0^2} \pmb{H}_n\left(k a_r\right) \right]\nonumber \\&\sim \frac{-2 {\gamma}_0^n}{\pi} \sin{(\pi \gamma_0)} \sum _{k=1}^\infty    \left[\frac{(-1)^k}{k^2} \pmb{H}_n\left(k a_r\right) \right]
     \label{ix1}
\end{align} 
\citet[Part II, Section 10, No. 36]{Mangulis2012}, where the approximation is valid for $|\gamma_0| \ll 1$.  From \citet[6.4.1-3]{Prudnikov}
\begin{eqnarray}  
\csc(\pi \gamma_0)\pmb{H}_1\left(\gamma_0  a_r\right) \sim \frac{2 {\gamma}_0}{3\pi^2} {a}_r^2
     \label{ix2}
\end{eqnarray} 
Comparing Eq.~(\ref{ix2}) to the $\beta-$term in (\ref{lam1}) gives
\begin{eqnarray}  
\frac{1}{a_r} \frac{d a_r}{ dt} \approx- \frac{\beta _2}{2}    \left(1+\frac{4}{3 \pi^2} a_r \right)  
     \label{i2}
\end{eqnarray} 
provided $|\sin(\pi \gamma_0)| \leq 1/(2  \beta)\ll 1$.   The solution to Eq.~(\ref{i2}) can be obtained by replacing  $g^*= 8 \beta |\gamma_0|/(3 \pi)$  in  Eqs.~(\ref{smlam2} and \ref{wrsm3})  by $4/(3 \pi^2)$ and is independent of $\gamma_0$. For TC1 with $\beta = 3.45$,  this is valid for $|\gamma_0| \leq 2.6^ \circ$ and for $|\gamma_0| \leq 6.5^ \circ$ for TC2 with $\beta = 1.43$. This interesting case has been experimentally demonstrated in figure 5(a) of \citet{kedr23} who showed the collapse of normalized responses of TC2 in terms of $\gamma^*$ for two different release angles of $\gamma_0 = -80 ^ \circ$ and $\gamma_0 = -40 ^ \circ$ only at large $t$ and very low $|\gamma^*|$, indicating  independence of $\gamma_0$. This also supports the use of low $\beta$ for wind vanes to extend the independence to a higher $\gamma_0$, with, however, the penalty of a slower decay rate.

The  analysis can be developed further to find the requirements for an exactly linear solution. The following relation holds:
\begin{eqnarray}  
\pmb{H}_1\left(\gamma_0  a_r\right)= \left((2\gamma_0)/\pi-d(\pmb{H}_0\left(\gamma_0  a_r\right))/d (a_r)\right)/\gamma_0.
     \label{i3}
\end{eqnarray} 
From Eq.~(\ref{ix1})
\begin{eqnarray}  
\pmb{H}_0\left(\gamma_0  a_r\right)= \frac{-2 }{\pi} \sin{(\pi \gamma_0)} \sum _{k=1}^\infty    \left[\frac{(-1)^k}{k^2-{\gamma}_0^2} \pmb{H}_0\left(k a_r\right) \right]\sim \frac{-2}{\pi} \sin{(\pi \gamma_0)} \sum _{k=1}^\infty    \left[\frac{(-1)^k}{k^2} \pmb{H}_0\left(k a_r\right) \right].\nonumber \\
     \label{i4}
\end{eqnarray} 
\citet[6.4.1-3]{Prudnikov} gives the sum of the last series as $-a_r/\pi$. Hence, $\pmb{H}_1\left(\gamma_0  a_r\right) \to 0$ in Eq.~(\ref{i3}) when
\begin{eqnarray}  
\sin{(\pi \gamma_0)\rightarrow \pi \gamma_0}.
     \label{i5}
\end{eqnarray} 
This makes the solution independent of $K_v$ at sufficiently small values of $\pi |\gamma_0|$ and leads also to $J_1(\gamma_0  a_r) \sim \left(\gamma_0  a_r \right)/2$. Using these relations, the solution becomes exactly linear, recovering the linear solution of $a_r$ and $\omega_r$ stated in Section \ref{sec:solution} and consequently giving the conditions for the validity of the small $|\gamma_0|$ linear solution in Eq.~(\ref{13ccc}).

We derived in this subsection  two important limiting cases of the current model whose validity depends on the magnitude of $\gamma_0$. The first makes the solution independent of $\beta$, and exactly linear: $\sin(\pi \gamma_0)/(\pi \gamma_0) \approx 1$. The second makes the solution independent of $\gamma_0$, but with small nonlinearity: $\sin(\pi \gamma_0) \le 1/(2\beta) \ll 1$. The two limits coincide as $\AR \rightarrow 0$, or equivalently $\beta \rightarrow \infty$ where the vortex flow dominates, \citet{devoria2017}. This condition may have led to the view that yaw response is linear for low $\AR$ but becomes nonlinear as $\AR$ increases but we have shown that this is true only for $\gamma_0 \rightarrow 0$. We have now reached the unexpected finding that  linearity requires the existence of vortex flow only, first for $\beta \rightarrow \infty$ as $|\gamma_0| \rightarrow 0$ and, second, as $\gamma_0 \rightarrow \pi/2$ for $\AR \rightarrow \infty$.

\subsection{Analytical solution for large yaw angle}
\label{sec:large}
To extend the linear solution from subsection \ref{bound}, we now derive the nonlinear  solution for  $|\gamma_0| \gg 1$. The asymptotic expansions of the $J_1(\gamma_0  a_r)$, $\pmb{H}_1\left(\gamma_0  a_r\right)$  from \citet[§9.2.1, and   §9.2.2 and §12.1.31, respectively]{abramowitz1972handbook},  are
\begin{equation}  
J_1(|\gamma_0|  a_r) \sim \sqrt{\frac{2}{\pi |\gamma_0|  a_r }} \left[\cos {\left(|\gamma_0|  a_r- \frac{3 \pi}{4}\right)}+\textit{O}\left(\frac{1}{|\gamma_0|  a_r}\right)\right] 
\label{la1}
\end{equation}
and
\begin{equation}  
\pmb{H}_1\left(\gamma_0  a_r\right) \sim \frac{2}{\pi}+\sqrt{\frac{2}{\pi |\gamma_0|  a_r }} \left[\sin {\left(|\gamma_0|  a_r- \frac{3 \pi}{4}\right)}+\textit{O}\left(\frac{1}{|\gamma_0|  a_r}\right)\right]. 
\label{la1x}
\end{equation}
Substituting $a^*_r= |\gamma_0| (a_r-1)$, Eq.~(\ref{lam1}) becomes 
\begin{eqnarray}  
 \frac{d a^*_r}{ dt} =-\beta_2  \left[\frac{4}{\pi} \beta+ \sqrt{\frac{2}{\pi |\gamma_0|}} \left\{\left(a^*_r-\frac{{a^*_r}^2}{2 |\gamma_0| } \right)\cos{\delta_a}+\left(1-\frac{a^*_r}{2 |\gamma_0| } \right)\sin{\delta_a} \right\}\right].  
     \label{la2}
\end{eqnarray}
where $\delta_a= |\gamma_0|-3 \pi/4+\phi_a$ and $\phi_a= \text{arccot}(2 \beta)\approx 1/(2 \beta)$. At small $t$, $a_r$ varies slowly so $a_r \approx 1$ and all the high order terms of $a^*_r$ can be safely neglected. This reduces the equation to
\begin{eqnarray}  
 \frac{d a^*_r}{ dt} =-N_1 a^*_r-M_1 .  
     \label{la3}
\end{eqnarray}
where
\begin{equation}  
 N_1= \beta_2\sqrt{\frac{2}{\pi |\gamma_0|}}\left[\cos{\delta_a}-\frac{\sin{\delta_a}}{2 |\gamma_0|}\right] \hspace{2mm}\text{and}\hspace{2mm}M_1=   
      \beta_2 \left[\frac{4}{\pi} \beta+ \sqrt{\frac{2}{\pi |\gamma_0|}}\sin{\delta_a}\right].   \label{la4}
\end{equation}
The solution of Eq.~(\ref{la3}) is
\begin{eqnarray}  
  a^*_r =\left[\left(M_1+N_1  {a^*_r}_0 \right) \exp{(-N_1 t)}-M_1\right]/N_1.  
     \label{la5}
\end{eqnarray} So, $a_r=a^*_r/|\gamma_0|+1$ and $ {a^*_r}_0=|\gamma_0|\left(a_{r0}-1\right)$.

At very small $t$, both $M_1$ and $N_1$ cancel from Eq.~(\ref{la5}) which makes the amplitude linear in $t$, but the exponential decay quickly becomes important.  This means the nonlinearity at high $\gamma_0$ results in more rapid decay  than a conventional linear system as shown by figure 5(a) of \citet{kedr25}.  At high $\gamma_0$, $M_1 \approx 4 \beta_2 \beta/\pi  \propto K_v$. Thus, having a tail fin with high $K_v$ is favourable in general. 

At small $|\gamma-\pi/2|$, the $K_p-$terms are negligible, as stated in Section \ref{bound}  and will be shown in Section \ref{sec:coeff}. This reduces $N_1$ and $M_1$ to
\begin{eqnarray}  
 N_1&=& 2 \beta \beta_2\sqrt{\frac{2}{\pi |\gamma_0|}}\left[\cos{\left( |\gamma_0|-\frac{3 \pi}{4}\right)}-\frac{\sin{( |\gamma_0|-3 \pi/4)}}{2 |\gamma_0|}\right]  
    , \text{and}\nonumber\\
    M_1&=&  2 \beta \beta_2 \left[\frac{2}{\pi}+ \sqrt{\frac{2}{\pi |\gamma_0|}}\sin{\left( |\gamma_0|-\frac{3 \pi}{4}\right)}\right]   \label{la6}
\end{eqnarray}
Similarly,
\begin{eqnarray}  
 \psi _r=\psi _{{r0}}-\frac{1}{N_1}\left[2(\sqrt{z}-\sqrt{{z}_i})+\Delta \log \left| \frac{\left(\sqrt{z}-\sqrt{\Delta}\right) \left(\sqrt{{z}_i}+\sqrt{\Delta}\right)}{\left(\sqrt{z}+\sqrt{\Delta}\right) \left(\sqrt{{z}_i}-\sqrt{\Delta}\right)}\right| \right]
  \label{la7}
\end{eqnarray}
where $\Delta= M_2-N_2M_1/N_1$, $z =M_2+N_2a^*_r$, and $z_i =M_2+N_2  {a^*_r}_0$.  $N_2$ and $M_2$ are given by
\begin{eqnarray}  
 N_2&=& 2 \beta_1 \left[-\frac{\beta}{\pi |\gamma_0|}+\frac{1}{\sqrt{\pi |\gamma_0|}} \left(\cos \delta_b-\frac{3}{4 |\gamma_0|}\sin \delta_b\right)\right], \text{and} \nonumber\\
    M_2&=&   \beta_1 \left[\frac{2\beta}{\pi}+\frac{1}{\sqrt{\pi |\gamma_0|}}\sin \delta_b\right] 
    \label{la9}
\end{eqnarray} and $\delta_b=2|\gamma_0|-3\pi/4+\phi_b$ and $\phi_b= \text{arccot}{\beta}\approx 1/\beta$.  Without the $K_p-$ terms:
\begin{eqnarray}  
 N_2&=& 2 \beta \beta_1 \left[-\frac{1}{\pi |\gamma_0|}+\frac{1}{\sqrt{\pi |\gamma_0|}} \left(\cos(2 |\gamma_0|-\frac{3 \pi}{4}\right)-\frac{3}{4 |\gamma_0|}\sin\left(2 |\gamma_0|-\frac{3 \pi}{4}\right)\right], ~~\text{and}\nonumber\\
    M_2&=& \beta  \beta_1 \left[\frac{2}{\pi}+\frac{1}{\sqrt{\pi |\gamma_0|}}\sin {\left(2 |\gamma_0|-\frac{3 \pi}{4}\right)}\right]. 
    \label{la10}
\end{eqnarray} 
The initial conditions for the TCs require
\begin{equation}  
 a_{r0}=\frac{{M}_1^2}{2({M}_2 |\gamma_0|-{M}_1{N}_1)|\gamma_0|}+1 \approx \frac{{M}_1^2}{2({M}_2 {\gamma}_0^2)}+1,\hspace{2mm}\text{and}\hspace{2mm}
    \psi _{r0}= \arccos(1/a_{r0}).
    \label{la11}
\end{equation} 
\subsubsection{Validation of the high angle solution}
The linear solution for $\gamma_0 \rightarrow \pi/2$, Eq.~(\ref{chi}), is useful in checking the range of application of the more general equation of the previous subsection.  Equation Eq.~(\ref{chi}) is accurate when $r_u \dot \gamma \ll 1$. At the beginning of motion where the phase angle $\psi_r \approx 0$ and the amplitude $a_r \approx 1$, it can be shown using  Eqs.~(\ref{6_4}, \ref{6_5}) and the definition of $\lambda$ in Eq.~(\ref{lam1}) that this term is of the order of $\textit{O}({\beta}_2^2)$ which is considered negligible in the current analysis.

The solution $\gamma= \gamma_0 a_r \cos{\psi_r}$  can be used for large $|\gamma|$ with $a_r$ and $\psi_r$ taken from Eqs.~(\ref{la5} and \ref{la7}) respectively. The series expansion of the solution after enforcing the initial conditions and neglecting the $\textit{O}({\beta}_2^2)$ terms reduces to : 
\begin{eqnarray}  
 \gamma= \gamma_0-\frac{1}{2} M_2 \gamma_0 t_*^2+\frac{1}{24} {M}_2^2 \gamma_0 t_*^4+ \textit{O}(t_*^6) .  
     \label{h2} 
\end{eqnarray}
where $t_*$ is defined in the first paragraph of Section 2.4.  The quadratic term $-1/2 M_2 \gamma_0 t_*^2$ has, from Eq.~(\ref{la10}) at $\gamma_0 = \pi/2$,  a factor of $3/\pi =0.955$ compared to the factor of unity in Eq.~(\ref{chi}).
In order to improve the accuracy of the high$-\gamma$ solution, the lowest order term omitted from  Eqs.~(\ref{la1} and \ref{la1x} ) is now added. From \citet{steinig1970}, the term is $3\sqrt{2}/\left(8\sqrt{\pi}) (|\gamma_0|  a_r)^{3/2}\right) \cos(|\gamma_0|  a_r-3 \pi/4)$. The new coefficients of Eqs.~(\ref{la6}, \ref{la10}) are 
\begin{eqnarray}  
 N_1&=& 2 \beta \beta_2\sqrt{\frac{2}{\pi |\gamma_0|}}\left[\left(1-\frac{9}{16 {\gamma}_0^2}\right)\cos{\left( |\gamma_0|-\frac{3 \pi}{4}\right)}-\frac{7}{8 |\gamma_0|}\sin\left( |\gamma_0|-\frac{3 \pi}{4}\right)\right]  
    , \nonumber\\
    M_1&=&  2 \beta \beta_2 \left[\frac{2}{\pi}+ \sqrt{\frac{2}{\pi |\gamma_0|}}\left\{\sin{\left( |\gamma_0|-\frac{3 \pi}{4}\right)}+\frac{3}{8 |\gamma_0|}\cos{\left( |\gamma_0|-\frac{3 \pi}{4}\right)}\right\}\right]   \label{g90_1}
\end{eqnarray}
and
\begin{eqnarray}  
 N_2&=& 2 \beta \beta_1 \left[-\frac{1}{\pi |\gamma_0|}+\frac{1}{\sqrt{\pi |\gamma_0|}} \left\{\left(1-\frac{15}{64 {\gamma}_0^2}\right)\cos{\left( 2|\gamma_0|-\frac{3 \pi}{4}\right)}-\frac{9}{16 |\gamma_0|}\sin\left( 2|\gamma_0|-\frac{3 \pi}{4}\right)\right\} \right] \nonumber\\
    M_2&=& \beta  \beta_1 \left[\frac{2}{\pi}+\frac{1}{\sqrt{\pi |\gamma_0|}}\left\{\sin {\left(2 |\gamma_0|-\frac{3 \pi}{4}\right)}+\frac{3}{16 |\gamma_0|}\cos{\left(2 |\gamma_0|-\frac{3 \pi}{4}\right)}\right\}\right].
    \label{g90_2}
\end{eqnarray} 
Now the approximation of $H_1(a_r \gamma_0)$ is accurate to $\textit{O}(|\gamma_0|^{-2})$.   For $\gamma_0=\pi/2$, $M_2$  has a factor of $ 3/\pi (1+ 1/(8\pi)) = 0.993$ compared to unity in Eq.~(\ref{chi}).

\section{Further developments of the minimal equation}
\label{sec:coeff}
Two important developments of the minimal  equation are discussed here. Neither of them affect the aim of the current study. The first is inclusion of  bearing friction. Yaw bearing friction for tail fins was  reviewed and studied by \citet{kedr25}.  Using system identification, they found that a static friction term, $k_s \sign (\dot \gamma)$, where $k_s$ is the Coulomb friction coefficient, was the most accurate simple model for their wind tunnel tests.   This term, which was also considered by \cite{BEECHAM1971}, can be simply added to Eq.~(\ref{3}). \cite{BEECHAM1971} point out that static friction in the TCs and similar situations should cause the fin to stick at $\gamma_{min}$, rather than reach $\gamma=0$.  Then $k_s=K_p  |\gamma|_{min}/I_*$ when $|\gamma|_{min}$ is sufficiently small to allow the assumption of Eq.~(\ref{i5}) which makes the solution  independent of $K_v$. figure 7 of \citet{kedr23}, however, shows the effect of friction at each peak of the response when $U$ has the low value of 5 m/s. Then, both $K_p$ and $K_v$ will contribute to $k_s$. This confirms the importance of model coefficient identification using wind tunnel tests  to describe the interaction between the aerodynamic coefficients and friction constants. \citet{kedr25} found that $\gamma_{min}$ was either constant or increases with $U^{-2}$ when it presumably depends mainly on the tail fin inertia which influences the yaw loads on the yaw bearings.  The ultimate conclusion is that friction will make the damping  dependent on $U$ and this would modify the current frictionless model. Fortunately, the solution of the static friction model is straightforward to accommodate within the current solution methods. 

The other issue is that Eq.~(\ref{3}) is, in effect, a quasi-steady equation as the only time dependent term in Eq.~(\ref{ang}) is generally small.  It is known, however, that the force coefficients, $K_p$ and $K_v$, for fins and vanes are not constant.  A simple allowance for their variation is through the separation functions ${x}_i^*$s in Eq.~(\ref{2}) which have been used in unsteady aerodynamics modelling for a long time; the most important references for  fins and vanes are given in \citet{HW23}. The functions describe the dynamic process of vortex motion over the fin surface including time lag and separation point movement. These functions are represented as first order differential equations with a time constant $\tau \propto c_0/U$. Because  fins and vanes typically have $c_0/x_p \ll 1$, their time scale is very small compared to the time scale of tail fin response $\sqrt{I_*}$. Thus,  \citet{HW23} assumed these functions depend only on $\gamma$ and not on $\dot{\gamma}$, in the simple form of $ {x}_{i,0}^*$:
\begin{equation}
   {x}_i^*\approx{x}_{i,0}^* (\gamma) =\left(1+ \exp [\sigma_i(|\gamma|-\alpha^{*}_i)]\right)^{-1} 
  \label{eq:56}
  \end{equation}
where $\sigma_i$ is an empirical constant that expresses the rate of decay of the forcing function and consequently the severity of stall. $\alpha^{*}_i$ expresses the angle shift in stall characteristics of the tail fin. \citet{kedr23} show that the term in $\dot{\gamma}$ which appears in the full form of Eq.~(\ref{eq:56}) was negligible for their wind tunnel tests.

$\alpha^{*}_i$ can have different values for each flow component. For potential flow in terms of $K_p$, it is the stall angle of the tail fin, and for  vortex flow in terms of $K_v$, the $\alpha$ at which the vortex  from the tail fin bursts at the apex. At this $\alpha$, the separation function ${x}_{i}^* = 0.5$. Both $\sigma_i$ and $\alpha^{*}_i$ are functions of the tail fin sweep angle, or alternatively $\AR$. 

These separation functions change the effective $K_p$ and $K_v$ in   Eqs.~(\ref{3} and \ref{4}) according to
\begin{equation}
      K_p \to {x}_{1,0}^* K_p, \hspace{2mm}K_v \to {x}_{2,0}^* K_v+(1-{x}_{3,0}^*) C_{D,c}
  \label{eq:57}
  \end{equation}
where $C_{D,c}$ is the drag coefficient of a normal flat plate which is function also of $\AR$; \citet{HW23}, however, argued that a representative value was $1.3$.   Thus, the static characteristics of the aerodynamic forces  change significantly with yaw angle rather than being described by constant coefficients. For example, the typical value of $K_v= \pi$  is reduced to $C_{D,c} = 1.3$ as $|\gamma| \to \pi/2$. In addition, ${x}_{1,0}^* \rightarrow 0$ as $|\gamma| \rightarrow \pi/2$ justifying the neglect of the $K_p$ term in this $\gamma$ range.  This neglect is used to simplify the limiting solutions for  $|\gamma_0|=\pi/2$ in Sections \ref{bound} and \ref{sec:large}.

A preliminary investigation showed the effect of separation functions on the analytic solution is, as expected, to change of the values of the coefficients $K_p$ and $K_v$ without affecting the form of the solution or its analysis.   In practice, an analytical solution including the separation functions is important for accurate prediction of the response compared to experimental measurements,  without changing the analysis given here. For that reason, the present study did not include the separation functions. 
\section{Conclusion}
\label{conc}
The current study describes approximate analytical solutions for the nonlinear yaw response of two important devices; tail fins for aligning small wind turbines with the wind, and wind vanes for  measurement of wind direction. Both are thin, swept surfaces that can have any planform and any release angle, $\gamma_0$.
The response is given in terms of damped second order differential equation of yaw angle, $\gamma$. The model is characterized by two main parameters:  $\beta_2 =r/U/\sqrt{I_*}$ and $\beta = K_v/K_p$, where $r$ is the moment arm length, $U$ is the wind velocity, which is assumed constant and coincident with the direction of the longitudinal inertial frame axis, and $\sqrt{I_*}$, the reduced inertia with dimension of s$^{-1}$.  This is the time scale of the current solution. $K_v$, $K_p$ are, respectively, the vortex flow and potential flow coefficients.  Both  are strong functions of the aspect ratio, $\AR$, which allows the application of the current models to any planform.  Before the present study, the only available analytical solution of the response equation was a second order, linear one involving $K_p$ only. It is restricted to small $|\gamma|$ and low $\AR$.  One of the main findings of this study is that $K_v$ influences the response at low $|\gamma|$ and Section 8.1 showed that including $K_v$ in the analysis of wind tunnel  results using system identification gave  values of $K_p$ and $K_v$  that are more consistent with those in the literature.  The literature values were usually determined from  steady experiments or calculations, justifying the restriction to a quasi-steady minimal equation of yaw response that is solved in the current study.

Our minimal equation has small nonlinear terms with two perturbation parameters; one for  $\gamma$ while the other for the yaw rate, $\dot \gamma$. Both are independent of $U$ but depend mainly on $\beta$; the latter depends also on $\beta_2$. They are linear in $\gamma_0$ at low  $|\gamma|$ and a function only of $K_v$ while becoming a complicated function of $\gamma_0$ at high $|\gamma|$. The small nonlinear damping terms makes the model soluble  with a perturbation method.  We used the averaging Krylov-Bogoliubov-Mitropolskii (KBM) method to obtain the amplitude $a$ and phase angle $\psi$ whose time derivative gives the theoretical frequency of the response. An analytical solution  is obtained by expanding the nonlinear function in a Chebyshev polynomial which is truncated to give a quintic polynomial of $a$ and approximate closed form solutions are obtained for $a$ and consequently $\psi$ that are valid for any $\gamma_0$. The  solution for $\gamma$ is a function of the model parameters and has a complicated dependence on $\gamma_0$.

The KBM solution is valid for ${\beta}_2^2 \ll1$ and found to be accurate compared to a numerical solution, but deviates a little as $|\gamma| \downarrow 0$ and the tail fin settles to equilibrium position at large $t$. Two test cases are used for the comparison; TC1 with low $\AR$ and high $\beta$, and TC2 with relatively high $\AR$ but low $\beta$. An unexpected finding was that the low $\beta$ (high $\AR$) causes smaller nonlinearity and lower error from truncating the  series used in finding the solution. This is in contrast to the previous linear theories which are linear \textit{because}  they are restricted to low $\AR$.  High $\AR$ gave, therefore, the more accurate response as $|\gamma| \downarrow 0$.  

The BT averaging method due to \citet{BEECHAM1971} was then used. It is more accurate in that it relaxed the first order approximation of yaw rate used in the KBM method and can accommodate the  ${\beta}_2^2$ terms in the solution.  Both of those, however, were found to be of marginal importance to the current study as most fins and vanes satisfy  ${\beta}_2^2 \ll 1$ and the KBM approximation to the yaw rate was found to be accurate for ${\epsilon} < 1$.  The BT method, however, provided compact expressions of two important solution parameters: the rate of amplitude decay, $(1/a) da/dt=\lambda$, and rate of change of phase angle, $d\psi/dt = {\omega}_r$. Neither have an analytical solution over all the $\gamma$ range except for the limiting cases, $|\gamma| \to 0$ and $|\gamma| \to \pi/2$. This large$-\gamma_0$ limit was introduced to allow a simple approximate analytic solution, Eq.~(\ref{chi}), but the general approximate solutions we derived are valid for any large value of $|\gamma_0|\ge \pi/2$.  The  limiting cases were used to define an equivalent linear system that has an equivalent damping ratio $\zeta_e=-\lambda/\omega_r$ and an equivalent frequency $\omega_r$. Further, an expression for the logarithmic decrement was obtained in terms of $\zeta_e$.

The compact expressions for $\lambda$ and $\omega_r$ enabled the derivation of closed form analytical solutions for the two limiting cases. For the low $|\gamma_0|$ solution, the amplitude  
decays at a rate that increases with $\beta$. Analysis of the low angle limiting solution shows that the response can restart from any extrema in $\gamma$ if  it is shifted in time from $\gamma_0$ and a closed form expression for this time shift is derived. 
The high angle limiting solution was also derived and validated against the exact analytical solution for $\gamma_0 =\pi/2$. The solution showed that the amplitude decay at very high $\gamma_0$ contains a linear and an exponential term indicating the relevance of high $K_v$ for tail fins which can experience high values of $|\gamma|$.

The region of linearity was found in terms of $\gamma_0$ to be wherever $\sin{(\pi \gamma_0)\approx \pi \gamma_0}$.  This  suggests the use of small $\AR$ planforms that have high $\beta$ for tail fins as the nonlinearity promotes faster response and rapid decay of $|\gamma|$.  In contrast, high $\AR$  for wind vanes will reduce their nonlinearity.  This suggestion is inconsistent with the linear, low$-\AR$ theory that is superseded here.

Finally, we note that our approximate analytic solutions to the minimal response equation were derived using the KBM and BT methods in a standard way. Nevertheless, several new mathematical results were required: for the Struve function in Eq.~(\ref{eqnH}), the Bessel function in Eq.~(\ref{eqnj2}) in terms of series of an odd Chebyshev polynomial with the former valid only for a positive argument of the Struve function. Also, section \ref{sec:epsilon} contains new series summations of the Weber function, for example, Eq.~(\ref{nnl7}). These may well have general application.
\begin{acknowledgments} The authors are grateful to Professor Francesco Castellani (Univ. Perugia, Italy) and Dr Amr Khedr (now at EPFL, Switzerland) for their major contributions to the experiments that are cited in the text and used to determine the  parameters of the minimal equation. The current study was unfunded. It grew out of a subcontract to the NREL "Distributed Wind Aeroelastic Modeling" program run by Brent Summerville (NREL, USA) to develop the model for a tail fin module for the aeroelastic software OpenFAST.  Dr Abhineet Gupta (NREL, USA) was instrumental in this work. 
\end{acknowledgments}

\appendix
\section{Derivation of the Chebyshev series of $f|f|$}\label{appA}

$f= \sin( \gamma_0 \gamma^*) + r_u\gamma_0 \dot{\gamma}^*$, so $f|f|$ is an odd function which when expanded in a  Fourier series gives
\begin{equation}
f|f|=-\frac{8}{\pi} \sum_{n=0}^{\infty} \frac{\sin\left( (2n+1) f\right) }{(2n+1)^3}+2\pi \sum_{n=0}^{\infty} (-1)^n \frac{\sin( (n+1) f) }{(n+1)}  .
  \label{A1}
\end{equation}
Let $f=\cos \theta$ and expand the $\sin$ terms in a Bessel series as:
\begin{eqnarray}
\sin({(2n+1)\cos \theta})&=& 2\sum_{m=0}^\infty (-1)^m J_{2m+1}(2n+1) \cos((2m+1) \theta)\nonumber\\ \sin({(n+1)\cos \theta})&=& 2\sum_{m=0}^\infty (-1)^m J_{2m+1}(n+1) \cos((2m+1)\theta).
  \label{A2}
\end{eqnarray}
Expressing $f|f|$ in terms of a Chebyshev series as
\begin{equation}
f|f|= \frac{b_0}{2}+\sum_{k=1}^\infty b_{k}T_{k}(f)
  \label{A3}
\end{equation}
where $f \in [-1,1]$. The  coefficients $b_{k}$ are determined as
\begin{equation}
b_{k}= \frac{2}{\pi}\int_0^\pi f|f| \cos(k\theta) d\theta.
  \label{A4}
\end{equation}
Substituting Eq.~(\ref{A2}) in (\ref{A1}) and the result in Eq.~(\ref{A4}), the integral in (\ref{A4}) is evaluated as
\begin{equation}
 \int_0^\pi  \cos((2m+1)\theta) \cos(k\theta) d\theta=  \frac{\pi}{2} \delta_{2m+1,k}
  \label{A5}
\end{equation}
where $\delta_{2m+1,k}$, the Dirac delta function, is zero for $k\neq 2m+1$.  Thus,  the Chebyshev series has only odd coefficients given by
\begin{equation}
 b_{2k+1}= -\frac{16}{\pi}  \sum_{n=0}^\infty \frac{ J_{2k+1}(2n+1)}{(2n+1)^3}+4\pi  \sum_{n=0}^\infty (-1)^n \frac{J_{2k+1}(n+1)}{(n+1)}  
  \label{A6}
\end{equation}
and $b_{2k}=0$. From \citet{Prudnikov} the following series summations are obtained:
\begin{eqnarray}
\sum_{n=0}^\infty \frac{ J_{2k+1}(2n+1)}{(2n+1)^3}&=& \frac{\pi^2}{16} \delta_{0,k}+\frac{1}{2(2k+3)(4k^2-1)}\nonumber\\ \sum_{n=0}^\infty (-1)^n\frac{J_{2k+1}(n+1)}{(n+1)}&=& -\sum_{n=1}^\infty (-1)^n \frac{J_{2k+1}(n)}{n}= 0  
  \label{A7}
\end{eqnarray}
where the second series summation is valid for $k>0$. For $k=0$ the second series gives
\begin{equation}
\sum_{n=1}^\infty (-1)^n \frac{J_{1}(n)}{n}= -\frac{1}{4}.
  \label{A8}
\end{equation}
This remaining term cancels with the $\delta_{0,k}$ term in the summation of the first series of Eq.~(\ref{A7}) resulting in
\begin{equation}
 b_{2k+1}= -\frac{8}{\pi} \frac{1} {(4 k^2-1)(2k+3)}, 
  \label{A9}
\end{equation}
so the Chebyshev series becomes
\begin{equation}
f|f|= -\frac{8}{\pi} \sum_{k=0}^\infty (-1)^k \frac{T_{2k+1}(f)} {(4 k^2-1)(2k+3)}.  
  \label{A10}
\end{equation}
\section {Simplification of the sign function terms in  Eq.~(\ref{4})}\label{appB}

The Chebyshev series for the sign term in Eq.~(\ref{3}) is 
\begin{eqnarray}
\left[\sin(\gamma) +r_u  \dot{\gamma}\right]^2 \sign{\left[\sin( \gamma)+r_u  \dot{\gamma}\right]}= -\frac{8}{\pi} \sum_{k=0}^{\infty}(-1)^{k}  \frac{T_{2k+1}(\sin {\gamma} +r_u \dot{\gamma})}{(4k^2-1)(2k+3)}  
  \label{appB:1}
\end{eqnarray}
After expanding $T_{2k+1}$ in a Taylor series to the second order in $r_u \dot{\gamma}$ and  some mathematical manipulation using Mathematica, we obtain
\begin{align}
 T_{2k+1}(\sin {\gamma} +r_u \dot{\gamma}) \approx&   (-1)^k [\cos((2k+1) \gamma)+ (2k+1) \frac{\cos((2k+1) \gamma)}{\cos \gamma}r_u\dot{\gamma}-\nonumber\\ &\frac{(2k+1)}{2 {\cos^3 \gamma} } \left((2k+1) \sin{2 k \gamma} +2k \sin{\gamma} \cos((2k+1) \gamma)\right){r}_u^2 \dot{\gamma}^2]+\textit{O}({r}_u^3\dot{\gamma}^3)
  \label{appB:2}
\end{align}
Substituting the Chebyshev expansion  into Eq.~(\ref{appB:1}) and summing the series using Mathematica gives the simplified term as
\begin{eqnarray}
\left[\sin(\gamma) +r_u  \dot{\gamma}\right]^2 \sign{\left[\sin( \gamma)+r_u  \dot{\gamma}\right]}\approx \sin \gamma |\sin \gamma|+ 2 r_u  |\sin \gamma|\dot{\gamma} + {r}_u^2 \dot{\gamma}^2\sign [\gamma]
  \label{appB:3}
\end{eqnarray}
to the third order of $\textit{O}({r}_u^3\dot{\gamma}^3)$, which is used in Eq.~(\ref{4}).
\section{Application of the KBM metohd}\label{appC}

This appendix describes the application of the KBM method as extended by \citet{Mendelson1970} for the general case of finite damping in Section \ref{sec:solution}.
We assume a periodic solution of Eq.~(\ref{6}) in the phase angle $\psi$ as
 \begin{equation}
  \gamma^*={\gamma}^{*}(a,\psi).
  \label{C2}
\end{equation} 
where the time-dependence of  $a$ and $\psi$ are  given  by:
\begin{equation}
  \frac{da}{dt}= \zeta(a),\hspace{2mm}\text{and}\hspace{2mm}\frac{d\psi}{dt}= \omega(a).
  \label{C3}
\end{equation}
$\gamma^*$, $\zeta$, and $\omega$ are expanded in power of the perturbation parameter $\epsilon$ as
\begin{eqnarray}
{\gamma}^*=& {\gamma}^{*}_{0}+ \epsilon {\gamma}^{*}_{1}+ {\epsilon}^2 {\gamma}^{*}_{2}+...\nonumber \\
\zeta=&-\sigma a+  \epsilon \zeta_1+ \epsilon^2 \zeta_2+...\nonumber \\
\omega=&\omega_0+  \epsilon \omega_1+ \epsilon^2 \omega_2+...
  \label{C4}
\end{eqnarray}
The leading terms in the expansions are chosen to give the linear solution of Eq.~(\ref{6}) as $\epsilon \to 0$.  Substituting the relations in  Eq.~(\ref{C4}) into Eq.~(\ref{6}) we obtain
\begin{eqnarray}
 \omega^2 \frac{\partial^2 {\gamma}^*}{\partial \psi^2}+ &&2 \omega \zeta \frac{\partial^2 {\gamma}^*}{\partial a \partial \psi}+\zeta^2 \frac{\partial^2 {\gamma}^*}{\partial a^2}+(\zeta \frac{d \omega}{da}+2 \sigma \omega)\frac{\partial {\gamma}^*}{\partial \psi}+(\zeta \frac{d \zeta}{da}+2 \sigma \zeta) \frac{\partial {\gamma}^*}{\partial a}+{\gamma}^* \nonumber\\&=& \epsilon F\left({\gamma}^*,\omega \frac{\partial {\gamma}^*}{\partial \psi}+\zeta \frac{\partial {\gamma}^*}{\partial a}\right) 
  \label{C5}
\end{eqnarray}
Equation (\ref{C5}) to zero order in $\epsilon$ is given by
\begin{equation}
 {\omega_0}^2 \frac{\partial^2 {\gamma}^{*}_{0}}{\partial \psi^2}- 2 \omega_0 \sigma a \frac{\partial^2 {\gamma}^{*}_{0}}{\partial a \partial \psi}+\sigma^2 a^2 \frac{\partial^2 {\gamma}^{*}_{0}}{\partial a^2}+2\sigma \omega_0\frac{\partial {\gamma}^{*}_{0}}{\partial \psi}-\sigma^2 a \frac{\partial {\gamma}^{*}_{0}}{\partial a}+({\omega_0}^2+\sigma^2){\gamma}^{*}_{0} =0.
  \label{C6}
\end{equation}
The solution to this equation is
\begin{equation}
 {\gamma}^{*}_{0}= a \cos\psi.
  \label{C7}
\end{equation}
Similarily, the first order equation  in $\epsilon$ is 
\begin{eqnarray}
 {\omega_0}^2 \frac{\partial^2 {\gamma}^{*}_{1}}{\partial \psi^2}&&- 2 \omega_0 \sigma a \frac{\partial^2 {\gamma}^{*}_{1}}{\partial a \partial \psi}+\sigma^2 a^2 \frac{\partial^2 {\gamma}^{*}_{1}}{\partial a^2}+2\sigma \omega_0\frac{\partial {\gamma}^{*}_{1}}{\partial \psi}-\sigma^2 a \frac{\partial {\gamma}^{*}_{1}}{\partial a}+({\omega_0}^2+\sigma^2){\gamma}^{*}_{1}\nonumber \\ &=&2\omega_0 \omega_1 a \cos \psi +2 \omega_0 \zeta_1 \sin \psi- \sigma a^2 \frac{d\omega_1}{da}\sin \psi+(\sigma a \frac{\zeta_1}{da}-\sigma\zeta_1) \cos \psi\nonumber \\&&+F(a \cos \psi, - \omega_0 a \sin \psi-\sigma a \cos \psi).
  \label{C8}
\end{eqnarray}
The function $F$ is periodic in $\psi$ as shown by  Eq.~(\ref{23}). The terms in $\cos \psi$ and $\sin \psi$ on the RHS of  Eq.~(\ref{C8}) must vanish to avoid secular terms in the solution. This condition leads to  Eqs.~(\ref{21} and \ref{22}). 

\section{Derivation of the Chebyshev series of $|\sin ({\gamma_0 \gamma^*})|$}\label{appD}
Expanding the even function $|\sin {\gamma_0 \gamma^*}|$ in a Fourier series gives
\begin{equation}
|\sin ({\gamma_0 \gamma^*})|= \frac{2}{\pi}-\frac{4}{\pi} \sum_{n=1}^\infty \frac{\cos(2n \gamma_0 \gamma^*)}{4n^2-1}.
  \label{d1}
\end{equation}
Let $\gamma^*=\cos \theta$ and substitute the  relation from \citet{Snyder}:
\begin{equation}
\cos({2n\gamma_0 \cos \theta})= J_0(2n\gamma_0)+2\sum_{m=1}^\infty (-1)^m J_{2m}(2n\gamma_0) \cos(2m \theta).
  \label{Aa2}
\end{equation}
This gives
\begin{equation}
|\sin ({\gamma_0 \cos \theta})|= \frac{2}{\pi}-\frac{4}{\pi} \sum_{n=1}^\infty \frac{J_0(2n\gamma_0)+2\sum_{m=1}^\infty (-1)^m J_{2m}(2n\gamma_0) \cos(2m \theta))}{4n^2-1}.
  \label{Aa3}
\end{equation}

Expressing $|\sin {\gamma_0 \gamma^*}|$ in terms of Chebyshev series as
\begin{equation}
|\sin ({\gamma_0 x})|= \frac{1}{2}b_0+\sum_{k=1}^\infty b_k T_k (\gamma^*)
  \label{Aa4}
\end{equation}
The series coefficients $b_k$ are determined as
\begin{equation}
b_k= \frac{2}{\pi}\int_0^\pi |\sin ({\gamma_0 \cos \theta})| \cos(k\theta) d\theta,
  \label{Aa5}
\end{equation}
or, from Eq.(\ref{Aa3})
\begin{eqnarray}
b_k&=& \frac{4}{\pi} \int_0^\pi  \cos(k\theta) d\theta-\nonumber \\&&\frac{8}{\pi} \int_0^\pi \sum_{n=1}^\infty \frac{J_0(2n\gamma_0)+2\sum_{m=1}^\infty (-1)^m J_{2m}(2n\gamma_0) \cos(2m \theta))}{4n^2-1} \cos(k\theta) d\theta 
  \label{Aa6}
\end{eqnarray}
Interchanging the order of summation and integration, the integral gives the coefficients as
\begin{equation}
b_k= \frac{4}{\pi} \left[\delta_{0,k}-2\delta_{0,k}\sum_{n=1}^{\infty} \frac{ J_0(2n\gamma_0)}{4n^2-1}\right]-\frac{8}{\pi}\sum_{k=1}^{\infty} (-1)^k  \left[\sum_{n=1}^{\infty} \frac{ J_{2k}(2n\gamma_0)}{4n^2-1}\right]
  \label{Aa7}
\end{equation}
where 
\begin{equation}
b_0= \frac{4}{\pi} \left[1-2\sum_{n=1}^{\infty} \frac{ J_0(2n\gamma_0)}{4n^2-1}\right].
  \label{Aa8}
\end{equation}
Substituting Eqs.~(\ref{Aa8} and \ref{Aa7}) into (\ref{Aa4}) and summing the series in (\ref{Aa7}) gives
\begin{equation}
  \sum_{n=1}^{\infty} \frac{ J_{2k}(2n|\gamma_0|)}{4n^2-1}= \frac{1}{4} \left[2\delta_{0,k}+\pi E_{2k}(|\gamma_0|)\right]
  \label{9}
\end{equation}
 where $\delta$ is the Dirac delta function which is zero for $k\neq 0$, e.g.  \cite{Prudnikov}. Substituting $k=0$ in Eq.~(\ref{9}) gives
\begin{equation}
  \sum_{n=1}^{\infty} \frac{ J_{0}(2n|\gamma_0|)}{4n^2-1}= \frac{1}{4} \left[2+\pi E_{0}(|\gamma_0|)\right].
  \label{10}
\end{equation}



\providecommand{\noopsort}[1]{}\providecommand{\singleletter}[1]{#1}%

\end{document}